\documentclass[english,aps,prb,twocolumn]{revtex4-1}
\usepackage[T1]{fontenc}
\usepackage[latin9]{inputenc}
\usepackage{color}
\usepackage{amsmath}
\usepackage{amssymb}
\usepackage{graphicx}
\usepackage{esint}

\makeatletter
\@ifundefined{textcolor}{}
{%
 \definecolor{BLACK}{gray}{0}
 \definecolor{WHITE}{gray}{1}
 \definecolor{RED}{rgb}{1,0,0}
 \definecolor{GREEN}{rgb}{0,1,0}
 \definecolor{BLUE}{rgb}{0,0,1}
 \definecolor{CYAN}{cmyk}{1,0,0,0}
 \definecolor{MAGENTA}{cmyk}{0,1,0,0}
 \definecolor{YELLOW}{cmyk}{0,0,1,0}
 }

\makeatother

\usepackage{babel}
\begin{document}

\newcommand{\1}{{\bf \scriptstyle 1}\!\!{1}}
\newcommand{\unit}{\overleftrightarrow{{\bf \scriptstyle 1}\!\!{1}}}
\newcommand{\I}{{\rm i}}
\newcommand{\p}{\partial}
\newcommand{\D}{^{\dagger}}
\newcommand{\hbe}{\hat{\bf e}}
\newcommand{\bfa}{{\bf a}}
\newcommand{\bx}{{\bf x}}
\newcommand{\hbx}{\hat{\bf x}}
\newcommand{\by}{{\bf y}}
\newcommand{\hby}{\hat{\bf y}}
\newcommand{\br}{{\bf r}}
\newcommand{\hbr}{\hat{\bf r}}
\newcommand{\bj}{{\bf j}}
\newcommand{\bk}{{\bf k}}
\newcommand{\bn}{{\bf n}}
\newcommand{\bv}{{\bf v}}
\newcommand{\bp}{{\bf p}}
\newcommand{\tp}{\tilde{p}}
\newcommand{\tbp}{\tilde{\bf p}}
\newcommand{\bu}{{\bf u}}
\newcommand{\hbz}{\hat{\bf z}}
\newcommand{\bA}{{\bf A}}
\newcommand{\calA}{\mathcal{A}}
\newcommand{\calB}{\mathcal{B}}
\newcommand{\tC}{\tilde{C}}
\newcommand{\bD}{{\bf D}}
\newcommand{\bE}{{\bf E}}
\newcommand{\calF}{\mathcal{F}}
\newcommand{\bB}{{\bf B}}
\newcommand{\bG}{{\bf G}}
\newcommand{\calG}{\mathcal{G}}
\newcommand{\obG}{\overleftrightarrow{\bf G}}
\newcommand{\bJ}{{\bf J}}
\newcommand{\bK}{{\bf K}}
\newcommand{\bL}{{\bf L}}
\newcommand{\tL}{\tilde{L}}
\newcommand{\bP}{{\bf P}}
\newcommand{\calP}{\mathcal{P}}
\newcommand{\bQ}{{\bf Q}}
\newcommand{\bR}{{\bf R}}
\newcommand{\bS}{{\bf S}}
\newcommand{\bH}{{\bf H}}
\newcommand{\balpha}{\mbox{\boldmath $\alpha$}}
\newcommand{\talpha}{\tilde{\alpha}}
\newcommand{\bsigma}{\mbox{\boldmath $\sigma$}}
\newcommand{\hbeta}{\hat{\mbox{\boldmath $\eta$}}}
\newcommand{\bSigma}{\mbox{\boldmath $\Sigma$}}
\newcommand{\bomega}{\mbox{\boldmath $\omega$}}
\newcommand{\bpi}{\mbox{\boldmath $\pi$}}
\newcommand{\bphi}{\mbox{\boldmath $\phi$}}
\newcommand{\hbphi}{\hat{\mbox{\boldmath $\phi$}}}
\newcommand{\btheta}{\mbox{\boldmath $\theta$}}
\newcommand{\hbtheta}{\hat{\mbox{\boldmath $\theta$}}}
\newcommand{\hbxi}{\hat{\mbox{\boldmath $\xi$}}}
\newcommand{\hbzeta}{\hat{\mbox{\boldmath $\zeta$}}}
\newcommand{\brho}{\mbox{\boldmath $\rho$}}
\newcommand{\bnabla}{\mbox{\boldmath $\nabla$}}
\newcommand{\bmu}{\mbox{\boldmath $\mu$}}
\newcommand{\bepsilon}{\mbox{\boldmath $\epsilon$}}

\newcommand{\iLambda}{{\it \Lambda}}
\newcommand{\cL}{{\cal L}}
\newcommand{\cH}{{\cal H}}
\newcommand{\cU}{{\cal U}}
\newcommand{\cT}{{\cal T}}

\newcommand{\be}{\begin{equation}}
\newcommand{\ee}{\end{equation}}
\newcommand{\bea}{\begin{eqnarray}}
\newcommand{\eea}{\end{eqnarray}}
\newcommand{\beqa}{\begin{eqnarray*}}
\newcommand{\eeqa}{\end{eqnarray*}}
\newcommand{\nn}{\nonumber}
\newcommand{\DD}{\displaystyle}

\newcommand{\ba}{\begin{array}{c}}
\newcommand{\baa}{\begin{array}{cc}}
\newcommand{\baaa}{\begin{array}{ccc}}
\newcommand{\baaaa}{\begin{array}{cccc}}
\newcommand{\ea}{\end{array}}

\newcommand{\bma}{\left[\begin{array}{c}}
\newcommand{\bmaa}{\left[\begin{array}{cc}}
\newcommand{\bmaaa}{\left[\begin{array}{ccc}}
\newcommand{\bmaaaa}{\left[\begin{array}{cccc}}
\newcommand{\ema}{\end{array}\right]}

\title{\textcolor{black}{Plasmonically enhanced mid-IR light source based on tunable spectrally and directionally selective thermal emission from nanopatterned graphene}}

\author{\textcolor{black}{Muhammad Waqas Shabbir$^{(1)}$}}

\author{\textcolor{black}{Michael N. Leuenberger$^{(1,2)}$}}

\email{michael.leuenberger@ucf.edu}

\affiliation{$^{(1)}$ NanoScience Technology Center and Department of Physics, University of Central Florida, Orlando, FL 32826, USA. \\
$^{(2)}$ College of Optics and Photonics, University of Central Florida, Orlando, FL 32826, USA.}

\begin{abstract}
We present a proof of concept for a spectrally selective thermal mid-IR source based on nanopatterned graphene (NPG) with a typical mobility of CVD-grown graphene (up to $3000$ cm$^2$V$^{-1}$s$^{-1}$), ensuring scalability to large areas. 
For that, we solve the electrostatic problem of a conducting hyperboloid with an elliptical wormhole in the presence of an {\it in-plane} electric field.
The localized surface plasmons (LSPs) on the NPG sheet allow for the control and tuning of the thermal emission spectrum in the wavelength regime from $\lambda =3$ $\mu$m to 12 $\mu$m by adjusting the size of and distance between the circular holes in a hexagonal or square lattice structure.
Most importantly, the LSPs along with an optical cavity increase the emittance of graphene from about 2.3\% for pristine graphene to 80\% for NPG, thereby outperforming state-of-the-art pristine graphene light sources operating in the near-infrared (NIR) by at least a factor of 100.
According to our COMSOL calculations, a maximum emission power per area of $11\times 10^3$ W/m$^2$ at $T=2000$ K for a bias voltage of $V=23$ V is achieved by controlling the temperature of the hot electrons through the Joule heating.
By generalizing Planck's theory and considering the nonlocal fluctuation-dissipation theorem with nonlocal response of surface plasmons in graphene in the random phase approximation (RPA), we show that the coherence length of the graphene plasmons and the thermally emitted photons can be as large as 13 $\mu$m and 150 $\mu$m, respectively, providing the opportunity to create phased arrays made of nanoantennas represented by the holes in NPG. The spatial phase variation of the coherence allows for beamsteering of the thermal emission in the range between $12^\circ$ and $80^\circ$ by tuning the Fermi energy between $\epsilon_F=-1.0$ eV and $\epsilon_F=-0.25$ eV through the gate voltage.  Our analysis of the nonlocal hydrodynamic response leads to the conjecture that the diffusion length and viscosity in graphene are frequency-dependent.
Using finite-difference time domain (FDTD) calculations, coupled mode theory, and RPA, we develop the model of a mid-IR light source based on NPG, which will pave the way to graphene-based optical mid-IR communication,  mid-IR color displays, mid-IR spectroscopy, and virus detection.

\textcolor{black}{KEYWORDS: Localized surface plasmons, graphene, emissivity, grey-body thermal emission, Planck's law, Stefan-Boltzmann law. }
\end{abstract}
\maketitle
An object that is kept in equilibrium at a given temperature $T>0$ K emits electromagnetic (EM) radiation because the charge carriers on the atomic and molecular scale oscillate due to their heat energy.\cite{Baranov2019} Planck's law describes quantitatively the energy density $u(\omega)$ of the EM radiation per unit frequency $\omega$ for black-body radiation, which is
$u_{BB}(\omega)d\omega=\frac{\omega^2}{\pi^2c^3}\Theta(\omega)d\omega$,
 where $c$ is the speed of light in vacuum, $\hbar$ is the Planck constant, and $k_B$ is the Boltzmann constant. $\Theta(\omega,T)=\hbar\omega/[\exp(\hbar\omega/k_BT)-1]$ is the thermal energy of a photon mode.
 Consequently, the energy emitted per unit surface area and per unit frequency, also called spectral radiance, of a black body into three-dimensional (3D) space is given by 
 \be
I_{BB}(\omega)d\omega=\frac{1}{4\pi}cu(\omega)=\frac{\omega^2}{4\pi^3c^2}\Theta(\omega)d\omega.
\label{eq:IBB}
\ee
 The total energy density $u$ can then be obtained by integrating over all frequencies and angles over the half-sphere, leading to the Stefan-Boltzmann law for the energy density of black-body radiation,
 \be
 u_{BB}=\left(\frac{8\pi^5k_B^4}{15c^3h^3}\right)T^4=a_{BB}T^4,
 \ee
 with $a_{BB}=7.566\times 10^{-16}$ Jm$^{-3}$K$^{-4}$.
 The total power emitted per unit surface area $P/A$ of a black-body is
 \bea
 I_{BB} & = &  \frac{P}{A}=\int\limits_0^\infty I_{BB}(\omega)d\omega \int\limits_0^{2\pi}d\varphi \int\limits_0^{\pi/2} \cos\theta\sin\theta d\theta \nn\\
 & = &  \pi\int\limits_0^\infty I_{BB}(\omega)d\omega = \frac{1}{4\pi}uc \nn\\
 & = & \frac{a_{BB}c}{4\pi}T^4=b_{BB} T^4=\left(\frac{\pi^2k_B^4}{60c^2\hbar^3}\right)T^4,
 \eea
 where $b_{BB}=5.67\times 10^{-8}$ Wm$^{-2}$K$^{-4}$ is the Stefan-Boltzmann constant.
 The factor $\cos\theta$ is due to the fact that black bodies are Lambertian radiators.

In recent years, several methods have been implemented for achieving a spectrally selective emittance, in particular narrowband emittance, which increases the coherence of the emitted  photons. One possibility is to use a material that exhibits optical resonances due to the band structure or due to confinement of the charge carriers.\cite{Baranov2019}
Another method is to use structural optical resonances to enhance and/or suppress the emittance. Recently, photonic crystal structures have been used to implement passive pass band filters that reflect the thermal emission at wavelengths that match the photonic bandgap.\cite{Cornelius1999,Lin2000} 
Alternatively, a truncated photonic crystal can be used to enhance the emittance at resonant frequencies.\cite{Celanovic2005,Yang2017}

Recent experiments have shown that it is possible to generate infrared (IR) emission by means of Joule heating created by means of a bias voltage applied to graphene on a SiO$_2$/Si substrate.\cite{Freitag2010,Luxmoore2013} In order to avoid the breakdown of the graphene sheet at around $T=700$ K, the graphene sheet can be encapsulated between hexagonal boron nitride (h-BN) layers, which remove efficiently the heat from graphene. The top layer protects it from oxidation.\cite{Kim2018,Luo2019} 
In this way, the graphene sheet can be heated up to $T=1600$ K,\cite{Luo2019} or even above $T=2000$ K.\cite{Kim2018,Shiue2019}
Kim et al. and Luo et al. demonstrated broadband visible emission peaked around a wavelength of $\lambda=725$ nm.\cite{Kim2018,Luo2019}
By using a photonic crystal substrate made of Si, Shiue et al. demonstrated narrowband near-IR emission peaked at around $\lambda=1600$ nm
with an emittance of around $\epsilon=0.07$.\cite{Shiue2019}
To the best of our knowledge, there are neither theoretical nor experimental studies on spectrally selective thermal emission from graphene in the mid-IR range.

Here, we present the proof of concept
of a method to tune the spectrally selective thermal emission from nanopatterned graphene (NPG) by means of a gate voltage that varies
the resonance wavelength of localized surface plasmons (LSPs) around the circular holes that are arranged in a hexagonal or square lattice pattern in a single graphene sheet
in the wavelength regime between 3 $\mu$m and 12 $\mu$m.
By generalizing Planck's radiation theory to grey-body emission, we show that the thermal emission spectrum can be tuned in or out of the two main atmospheric transparency windows of 3 to 5 $\mu$m and 8 to 12 $\mu$m in the mid-IR regime, and also in or out of the opaque mid-IR regime between 5 and 8 $\mu$m.
In addition, the gate voltage can be used to tune the direction of the thermal emission due to the coherence between the localized surface plasmons (LSPs) around the holes due to the nonlocal response function in graphene, which we show by means of a nonlocal fluctuation-dissipation theorem.
The main element of the nanostructure is a circular hole of diameter $a$ in a graphene sheet.
Therefore let us focus first on the optoelectronic properties of a single hole.

\begin{figure}
\begin{centering}
\includegraphics[width=8.5cm]{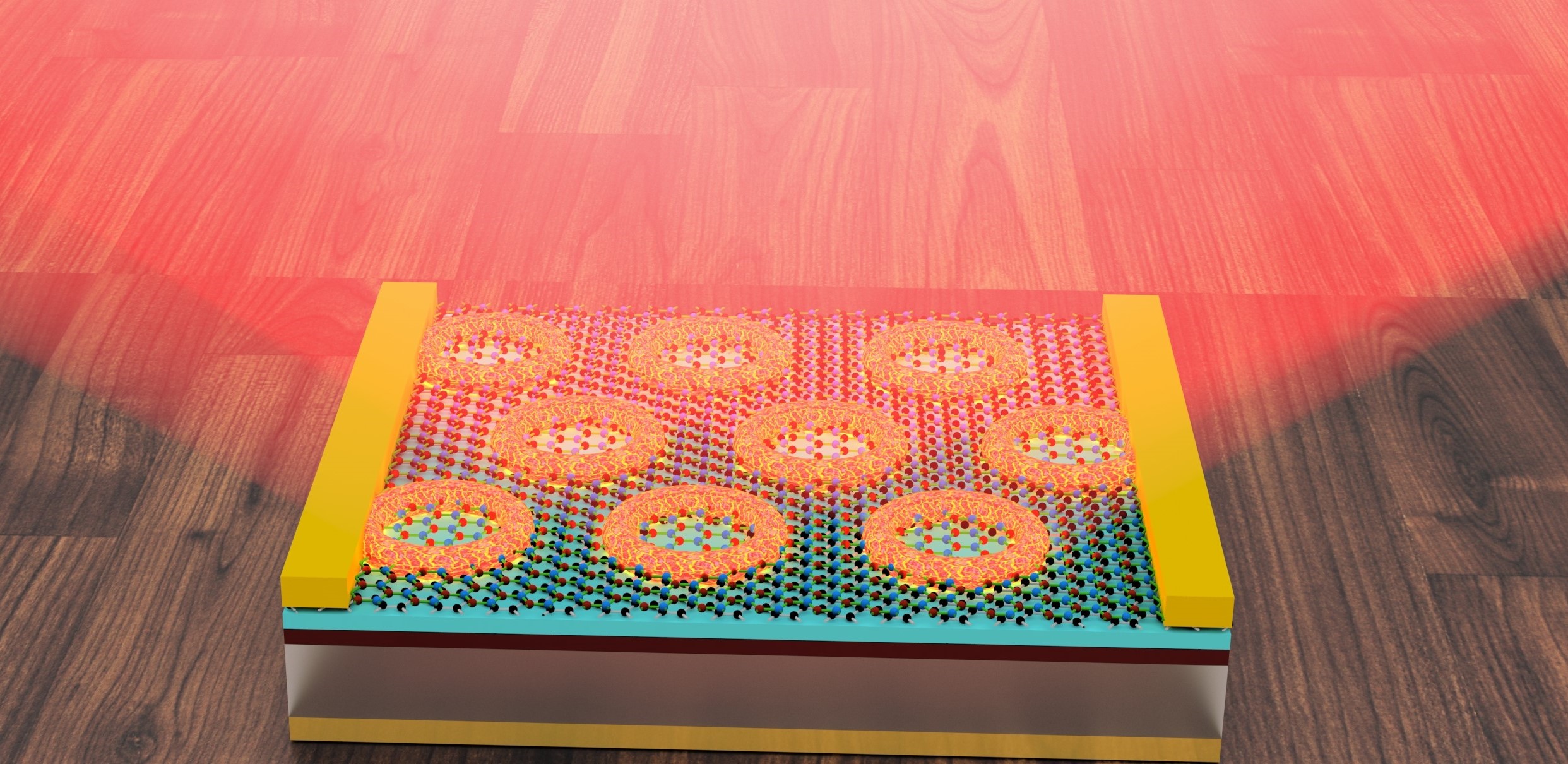}
\end{centering}
\caption{Schematic showing our proposed ultrafast mid-IR light source based on patterned graphene placed on top of a cavity, which can be tuned by means of a gate voltage applied to the ITO layer.\label{fig:graphene_emitter} }
\end{figure}

The frequency-dependent dipole moment of the hole is 
\bea
\bp(\br,\omega) & = & -\varepsilon_0\varepsilon(\br,\omega)\bE_{0||} \nn\\
& = & -\alpha_{1,2}(\br,\omega)\bE_{0||},
\eea 
where the polarizabilities $\alpha_{1,2}$ are given along the main axes $x$ and $y$ of the elliptic hole,
and $\br=\br_0$ is the position of the dipole moment, i.e. the hole.
Graphene's dielectric function is isotropic in the xy-plane, i.e. $\varepsilon''_{||}=\varepsilon''_{xx}=\varepsilon''_{yy}$.
$V_0$ is the volume of the graphene sheet.
In the Supplementary Information we derive the general polarizabilities of an uncharged single-sheet hyperboloid with dielectric function $\varepsilon(\omega)$ inside a medium with dielectric constant $\varepsilon_m$ [see Eq.~(\ref{eq:polarizability_hyperboloidx})].
The polarizabilities of an elliptical wormhole in $x$- and $y$-direction read
\bea
\alpha_1(\omega) & = & \frac{2abd\pi(\pi/2-1)}{3}\frac{\varepsilon(\omega)-\varepsilon_m}{\varepsilon_m+L_1[\varepsilon(\omega)-\varepsilon_m]}, \\
\alpha_2(\omega) & = & \frac{2abd\pi(\pi/2-1)}{3}\frac{\varepsilon(\omega)-\varepsilon_m}{\varepsilon_m+L_2[\varepsilon(\omega)-\varepsilon_m]}.
\label{eq:polarizability_hyperboloid_main}
\eea
respectively, for which the in-plane polarizabilities lies in the plane of the graphene sheet that is parallel to the $xy$-plane.
$\varepsilon(\omega)$ is the dielectric function of graphene.
We assumed that the thickness $d$ of the graphene sheet is much smaller than the size of the elliptic hole.
The geometrical factors in this limit are
\bea
L_1 & \approx & abd\int\limits_{\eta_1}^\infty \frac{d\eta'}{(\eta'+a^2)R_{\eta'}}, \\
L_2 & \approx & abd\int\limits_{\eta_1}^\infty \frac{d\eta'}{(\eta'+b^2)R_{\eta'}}.
\eea
In the case of a circular hole of diameter $a$ the polarizability simplifies to
\be
\alpha_{||}(\omega) = \frac{2a^2d\pi(\pi/2-1)}{3}\frac{\varepsilon(\omega)-\varepsilon_m}{\varepsilon_m+L_{||}[\varepsilon(\omega)-\varepsilon_m]}, 
\label{eq:polarizability_hyperboloid_circular}
\ee
The localized surface plasmon resonance (LSP) frequency of the hole can be determined from the equation
\be
\varepsilon_m+L_{||}[\varepsilon(\omega)-\varepsilon_m]=0,
\label{eq:plasmon}
\ee
the condition for which the denominator of $\alpha_{||}$ vanishes.

\begin{figure}[htb]
\begin{centering}
\includegraphics[width=7.5cm]{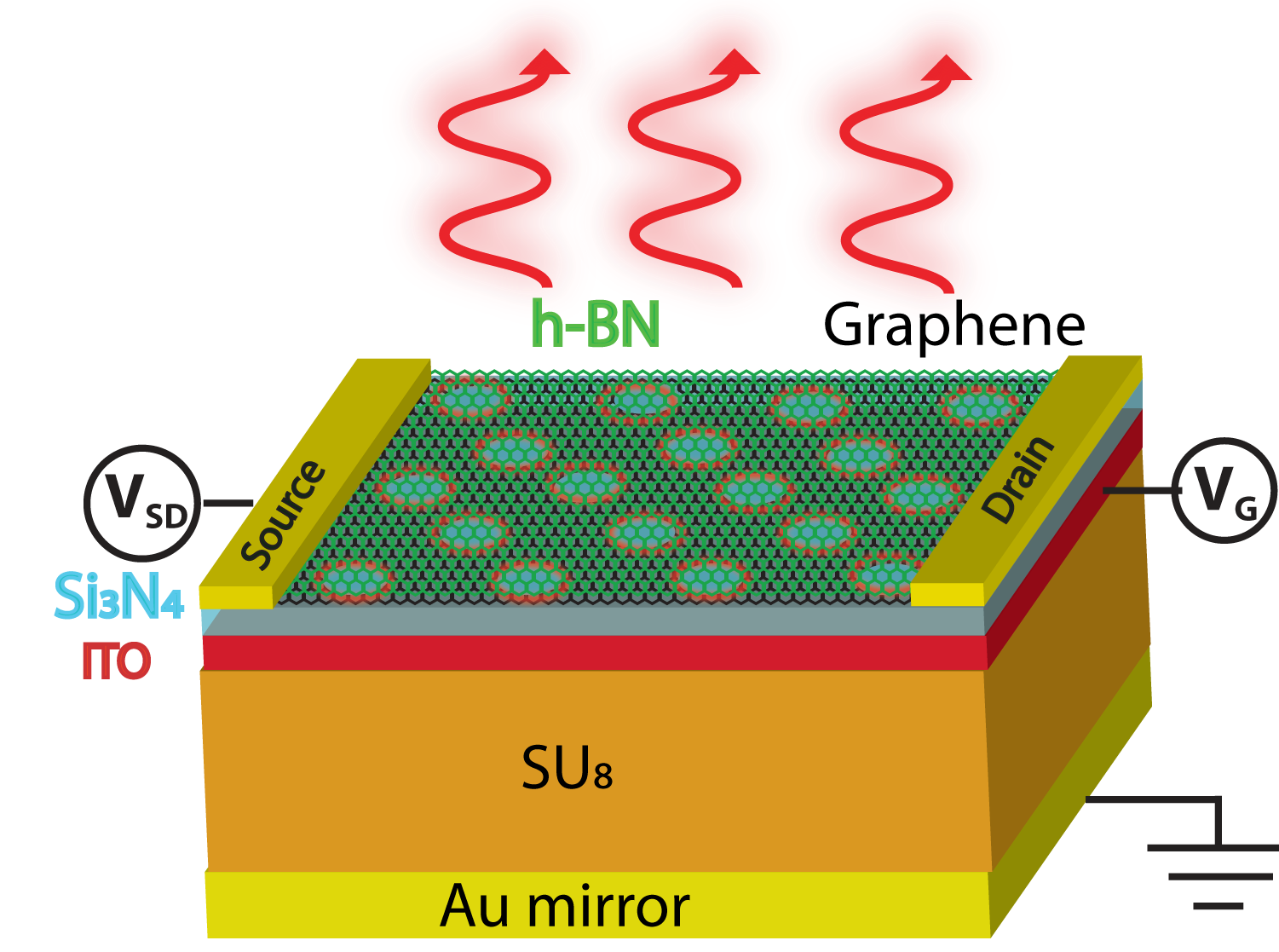}
\end{centering}
\caption{Schematic showing our proposed ultrafast mid-IR light source with the materials used in our setup.
The materials from top to bottom are:
1 single layer of hexagonal boron nitride (h-BN), for preventing oxidation of graphene at higher temperatures,
1 single layer of patterned graphene,
50 nm of Si$_3$N$_4$,  for large n-doping and gating,
50 nm of ITO, metallic contact for gating, which is also transparent in mid-IR,
$\lambda/4n_{\rm SU8}$ of SU$_8$,\cite{Safaei2017} which is transparent in mid-IR, and
Au back mirror. $n_{\rm SU8}=1.56$ is the refractive index of SU$_8$.
\label{fig:graphene_emitter_materials} }
\end{figure}

Using the linear dispersion relation, the intraband optical conductivity is\cite{Safaei2017,Paudel2017}
\be
\sigma _{\rm intra}(\omega ) = \frac{e^2}{\pi\hbar^2}\frac{2k_BT}{\tau^{-1} - i\omega}\ln \left[ 2\cosh \left( \frac{\varepsilon _F}{2k_BT} \right) \right],
\ee
 which in the case of ${\varepsilon _F} \gg {k_B}T$   is reduced to 
 \be
\sigma_{\rm intra}(\omega) = \frac{e^2}{\pi\hbar^2}\frac{\epsilon_F}{\tau^{-1} - i\omega }=\frac{2\varepsilon_m\omega_p^2}{\pi\hbar^2(\tau^{-1}-i\omega)},
\label{eq:sigma_intra}
 \ee
 where $\tau$ is determined by impurity scattering and electron-phonon interaction ${\tau ^{ - 1}} = \tau _{imp}^{ - 1} + \tau _{e - ph}^{ - 1}$ .
 Using the mobility $\mu$ of the NPG sheet, it can be presented in the form
 $\tau^{-1}=ev_F^2/(\mu E_F)$, where $v_F=10^6$ m/s is the Fermi velocity in graphene.
 $\omega_p=\sqrt{e^2\epsilon_F/2\varepsilon_m}$ is the bulk graphene plasma frequency.
 
 It is well-known by now that hydrodynamic effects play an important role in graphene because the Coulomb interaction collision rate is dominant, 
 i.e. $\tau _{ee}^{ - 1}\gg\tau _{imp}^{ - 1}$ and $\tau _{ee}^{ - 1}\gg\tau _{e - ph}^{ - 1}$, which corresponds to the hydrodynamic regime.
 $\tau _{imp}^{ - 1}$ and $\tau _{e - ph}^{ - 1}$ are the electron-impurity and electron-phonon collision rates.
 Since for large absorbance and emittance, we choose a large Fermi energy, we are in the Fermi liquid regime of the graphene sheet.
 Taking the hydrodynamic correction into account, we also consider the hydrodynamically adjusted intraband optical conductivity,\cite{Bandurin2016,Christensen2017}
  \be
\sigma_{\rm intra}^{\rm HD}(\omega)=\frac{\sigma_{\rm intra}(\omega)}{1-\eta^2\frac{k_{||}^2}{\omega^2}},
\label{eq:sigma_HD}
 \ee
 where $\eta^2=\beta^2+D^2\omega(\gamma+i\omega)$, $\beta^2\approx \frac{3}{4}v_F^2$ is the intraband pressure velocity, $D\approx 0.4$ $\mu$m is the diffusion length in graphene, and $\gamma=\tau^{-1}$ is the relaxation rate.
 Interestingly, the optical conductivity becomes $k$-dependent and nonlocal. Also, below we will conjecture that the diffusion length $D$ must be frequency-dependent.
 
 Note that since $\varepsilon=1+\chi$, where $\chi$ is the susceptibility, it is possible to replace $\varepsilon''=\chi''$.
Alternatively, using the formula of the polarizability $\alpha=\varepsilon_0\chi$ we can write $\varepsilon''=\alpha''/\varepsilon_0$.
The dielectric function for graphene is given by\cite{Safaei2017,Paudel2017}
\be
\varepsilon_{||}(\omega)=\varepsilon_g-\frac{i\sigma_{2D}(\omega)}{\varepsilon_0\omega d},
\ee
where $\epsilon_g=2.5$ is the dielectric constant of graphite and $d$ is the thickness of graphene.
Inserting this formula into Eq.~(\ref{eq:plasmon}) gives
\be
\varepsilon_m+L_{||}[\varepsilon_g-i \frac{e^2}{\pi\hbar^2}\frac{\epsilon_F}{\tau^{-1}\varepsilon_0\omega d}-\frac{i}{\varepsilon_0 d}-\varepsilon_m]=0,
\ee
Solving for the frequency and using the real part we obtain the LSP frequency,
\be
{\rm Re}\omega_{\rm LSP} = \frac{{2L_{||}^2\varepsilon_m\omega_p^2\tau}}{{\pi\hbar^2 \left\{ {{L^2} + d^2\varepsilon_0^2{{\left[ {L_{||}\left( {\varepsilon_g - \varepsilon_m} \right) +\varepsilon_m} \right]}^2}} \right\}}},
\ee
which is linear in the Fermi energy $\epsilon_F$.

\begin{figure}[htb]
\begin{centering}
\includegraphics[width=9.5cm]{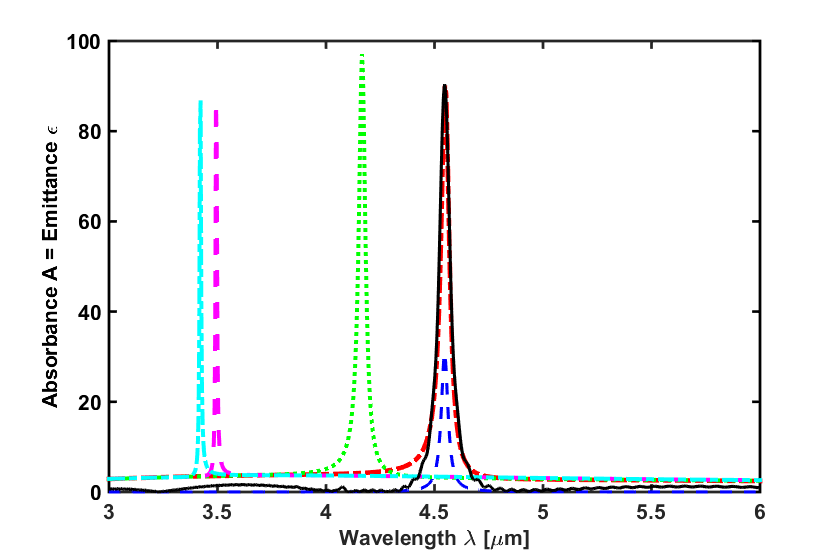}
\end{centering}
\caption{Emittance $\epsilon(\lambda)$ [equal to absorbance $A(\lambda)$] of the structure shown in Figs.~\ref{fig:graphene_emitter} and \ref{fig:graphene_emitter_materials} with Fermi energy $E_F=-1.0$ eV, mobility $\mu=3000$ V/cm$^2$s, hole diameter of $a=30$ nm, and period of $\calP=45$ nm  at $T=300$ K. The solid (black) curve represents the result of FDTD calculation.  The dashed (blue) curve and the solid (black) curve are the emittances $\epsilon_g$ and $\epsilon_{\rm FP}$ calculated by means of Eq.~(\ref{eq:emittanceg}) and  Eq.~(\ref{eq:emittance}) for the bare NPG sheet and the whole structure including cavity, respectively. 
The dotted (green) line exhibits a blue-shift due to the hydrodynamic correction shown in Eq~(\ref{eq:sigma_HD}) with $D(\nu=30$ THz$)\approx 0$.
The blue-shifted dashed (magenta) curve and the  blue-shifted dot-dashed (cyan) curve are the RPA-corrected LSP peaks due to the Coulomb interaction and the Coulomb interaction including electron-phonon interaction with the optical phonons of graphene, boron nitride, and Si$_3$N$_4$. This NPG sheet emits into the atmospheric transparency window between $3$ and $5$ $\mu$m.
\label{fig:emittance_4m} }
\end{figure}

Let us now consider the 2D array of circular holes in a graphene sheet.
Since the dipole moments $p_j=\delta\bp(\bR_j,\omega)$ interact with each other by inducing dipole moments, we need to consider the dressed dipole moment
at each site $\bR_j$ as source of the electric field, which is
\be
\tp_j =p_j+ \alpha\sum\limits_{j'\ne j}\calG_{jj'}\tp_{j'},
\ee
where $\calG_{jj'}$ is the dipole-dipole interaction tensor.
Using Bloch's theorem $p_j=p_0\exp(i\bk_{||}\cdot\bR_{||})$, the effective dipole moment becomes
\be
\tp_0 =p_0+\tp_0\alpha\sum\limits_{j'\ne j}\calG_{jj'}e^{i\bk_{||}\cdot(\bR_j-\bR_{j'})}.
\ee
for each site $j$, and thus
\be
\tp_0=\frac{p_0}{1-\alpha\calG}.
\ee
The lattice some over the dipole-dipole interaction tensor $\calG=\sum\limits_{j'\ne j}\calG_{jj'}e^{i\bk_{||}\cdot(\bR_j-\bR_{j'})}$ can be found in Ref.~\onlinecite{Thongrattanasiri2012}, i.e.
\bea
{\rm Re}\calG & \approx & g/\calP^3 , \\
{\rm Im}\calG & = &  S-2k^3/3,
\eea
where $\calP$ is the lattice period,
\be
S=\frac{2\pi k}{\Omega_0}\times\left\{\baa
\arccos\theta & \mbox{ for s polarization, } \\
\cos\theta & \mbox{ for p polarization. }
\ea
\right.,
\ee
$\Omega_0$ is the unit-cell area, and the real part is valid for periods much smaller than the wavelength.
The factor $g=5.52$ ($g=4.52$) for hexagonal (square) lattice.
The electric field created by the effective dipole moment is determined by
\be
\tbp_0=\talpha\bE,
\ee
from which we obtain the effective polarizability of a hole in the coupled dipole approximation (CDA),
\be
\talpha=\frac{\alpha}{1-\alpha\calG}.
\ee
This formula is the same as in Refs.~\onlinecite{Zhao2003,Thongrattanasiri2012}, where the absorption of electromagnetic waves by arrays of dipole moments and graphene disks were considered, respectively.
Thus, our result corroborates Kirchhoff's law (see below).
Consequently, we obtain the same reflection and transmission amplitudes as in Ref.~\onlinecite{Thongrattanasiri2012}, i.e.
\be
r = \frac{\pm iS}{\alpha^{-1}-\calG}, \; t = 1+r,
\ee
where the upper (lower) sign and $S=2\pi\omega/c\Omega_0\cos\theta$ ($S=2\pi\omega\cos\theta/c\Omega_0$) apply to s (p) polarization.
Thus, the emittance and absorbance of the bare NPG sheet are given by
\be
\epsilon_g=A_g=1-|r|^2-|t|^2.
\label{eq:emittanceg}
\ee
The coupling to the interface of the substrate with reflection and transmission amplitudes $r_0$ and $t_0$, respectively, which is located basically at the same position as the NPG sheet,
yields the combined reflection and transmission amplitudes\cite{Thongrattanasiri2012}
\be
R=r+\frac{tt'r_0}{1-r_0r'}, \; T=\frac{tt_0}{1-r_0r'},
\ee
where $r'=r$ and $t'=1-r$ are the reflection and transmission amplitudes in backwards direction, respectively.

\begin{figure}[htb]
\begin{centering}
\includegraphics[width=9.5cm]{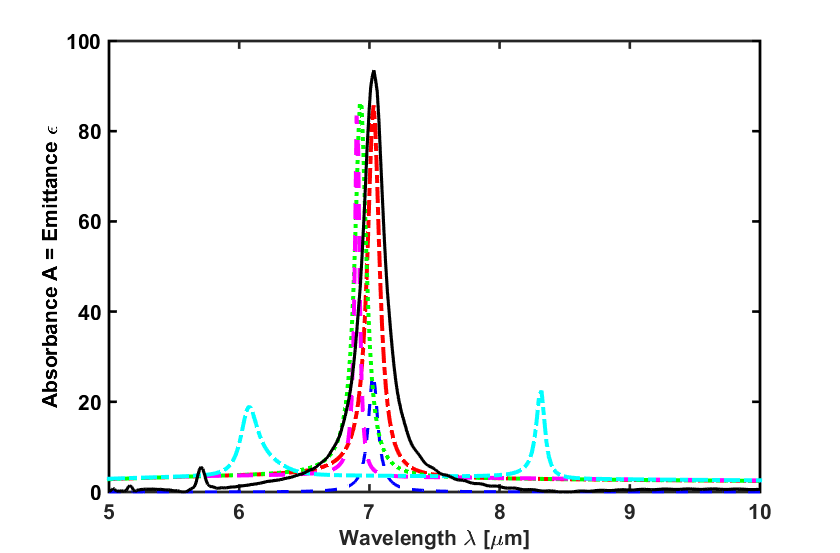}
\end{centering}
\caption{Emittance $\epsilon(\lambda)$ [equal to absorbance $A(\lambda)$] of the structure shown in Figs.~\ref{fig:graphene_emitter} and \ref{fig:graphene_emitter_materials} with Fermi energy $E_F=-1.0$ eV, mobility $\mu=3000$ V/cm$^2$s, hole diameter of $a=90$ nm, and period of $\calP=150$ nm  at $T=300$ K. The curves are denoted the same as in Fig.~\ref{fig:emittance_4m}. This NPG sheet emits into the atmospheric opacity window between $5$ and $8$ $\mu$m.
\label{fig:emittance_7m} }
\end{figure}

If we include also the whole substrate including cavity and Au mirror, we need to sum over all possible optical paths in the Fabry-Perot cavity, yielding
\be
R_{\rm FP}=R+TT'r_{\rm Au}e^{i\delta}\sum\limits_{m=0}^\infty r_m^m,
\ee
with
\be
r_m= r_{\rm Au}R'e^{i\delta},
\ee
where $r_{Au}$ is the complex reflection amplitude of the Au mirror in the IR regime.
$\delta=2kL\cos\theta$ is the phase accumulated by one back-and-forth scattering inside the Fabry-Perot cavity of length $L$.
$k\approx n_{SU_8}k_0$ is the wavenumber inside the cavity for an external EM wave with wavenumber $k_0=2\pi/\lambda$.
Since the sum is taken over a geometric series, we obtain
\be
R_{\rm FP}=R+\frac{TT'r_{\rm Au}e^{i\delta}}{1- r_{\rm Au}R'e^{i\delta}}.
\ee
Since the transmission coefficient through the Au mirror can be neglected, we obtain the emittance $\epsilon$ and absorbance $A$ including cavity, i.e.
\be
\epsilon_{\rm FP}=A_{\rm FP}=1-|R_{\rm FP}|^2.
\label{eq:emittance}
\ee

\begin{figure}[htb]
\begin{centering}
\includegraphics[width=9.5cm]{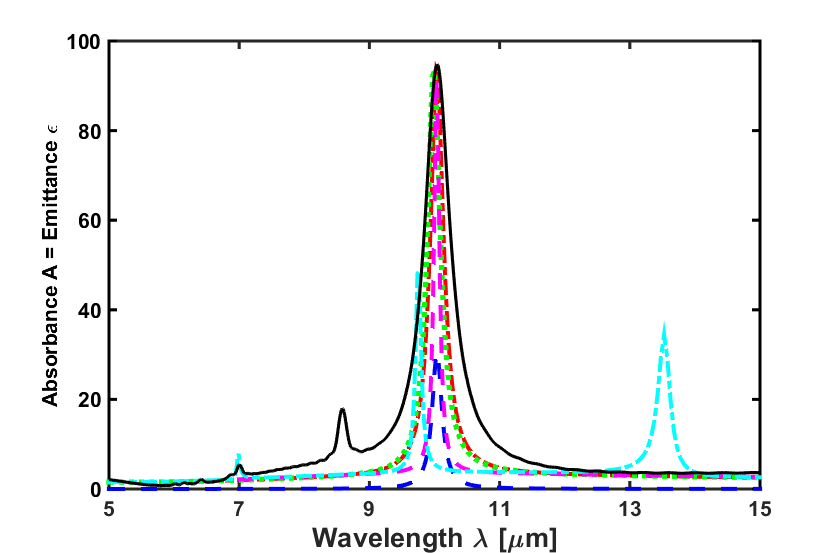}
\end{centering}
\caption{Emittance $\epsilon(\lambda)$ [equal to absorbance $A(\lambda)$] of the structure shown in Figs.~\ref{fig:graphene_emitter} and \ref{fig:graphene_emitter_materials} with Fermi energy $E_F=-1.0$ eV, mobility $\mu=3000$ V/cm$^2$s, hole diameter of $a=300$ nm, and period of $\calP=450$ nm at $T=300$ K. The curves are denoted the same as in Fig.~\ref{fig:emittance_4m}. This NPG sheet emits into the atmospheric transparency window between $8$ and $12$ $\mu$m.
\label{fig:emittance_10m} }
\end{figure}

Using these results, let us consider the excitation of the graphene sheet near the hole by means of thermal fluctuations, which give rise to a fluctuating EM field of a localized surface plasmon (LSP).
This can be best understood by means of the fluctuation-dissipation theorem, which provides a relation between the rate of energy dissipation in a non-equilibrium system
and the quantum and thermal fluctuations occuring spontaneously at different times in an equilibrium system.\cite{Novotny2012}
The standard (local) fluctuation-dissipation theorem for fluctuating currents $\delta\hat{J}_\nu(\br,\omega)$ in three dimensions reads
\bea
\left<\delta\hat{J}_\mu(\br,\omega)\delta\hat{J}_\nu(\br',\omega')\right> & = & \omega\varepsilon_0\varepsilon''_{\mu\nu}(\br,\omega)
\Theta(\omega) \nn\\
& & \times\delta(\omega-\omega')\delta(\br-\br'),
\eea
where the relative permittivity $\varepsilon(\br,\omega)=\varepsilon'(\br,\omega)+i\varepsilon''(\br,\omega)=f(\br)\varepsilon(\omega)$ and $\mu,\nu=x,y,z$ are the coordinates.
Note that since $\varepsilon=1+\chi$, where $\chi$ is the susceptibility, it is possible to replace $\varepsilon''=\chi''$.
Alternatively, using the formula of the polarizability $\alpha=\varepsilon_0\chi$ we can write $\varepsilon''=\alpha''/\varepsilon_0$.
$f(\br)=1$ on the graphene sheet and 0 otherwise.
Since the fluctuating currents are contained inside the two-dimensional graphene sheet, 
we write the local fluctuation-dissipation theorem in its two-dimensional form, i.e.
\bea
\left<\delta\hat{J}_\mu(\br_{||},\omega)\delta\hat{J}_\nu(\br_{||}',\omega')\right> & = & {\sigma'}_{\mu\nu}^{2D}(\br_{||},\omega)
\Theta(\omega) \nn\\
& & \times\delta(\omega-\omega')\delta(\br_{||}-\br_{||}'),
\label{eq:fluctuation-disspation_2D}
\eea
where the fluctuating current densities have units of A/m$^2$ and the coordinates are in-plane of the graphene sheet.

Using the method of dyadic Green's functions, it is possible to express the fluctuating electric field generated by the fluctuating current density by
\be
\delta\hat{\bE}(\br,\omega)=i\omega\mu_0\int_{\Omega}\bG(\br,\br_{0||};\omega)\delta\hat{\bJ}(\br_{0||},\omega)d^2r_{0||},
\ee
where $\Omega$ is the surface of the graphene sheet.
The LSP excitation around a hole can be well approximated by a dipole field such that 
\bea
\delta\hat{\bJ}(\br_{0||},\omega) & = & -i\omega\sum\limits_j\delta\tbp(\bR_j,\omega) \nn\\
& = & -i\omega\delta\tbp_0(\omega)\sum\limits_{j}\delta(\br_{0||}-\bR_j) ,
\eea
where $\bR_j=(x_j,y_j)$ are the positions of the holes in the graphene sheet.

\begin{figure}[htb]
\begin{centering}
\includegraphics[width=8.5cm]{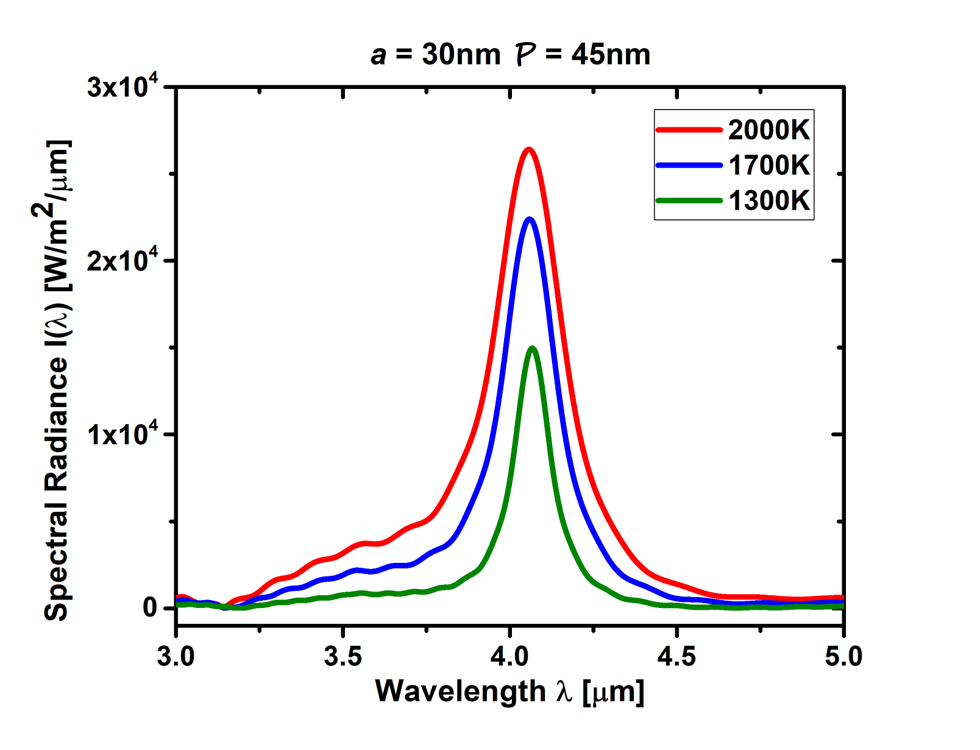}
\end{centering}
\caption{Spectral radiance of NPG including cavity, as shown in in Figs.~\ref{fig:graphene_emitter} and \ref{fig:graphene_emitter_materials},
 as a function of wavelength $\lambda$
with Fermi energy $E_F=-1.0$ eV, mobility $\mu=3000$ V/cm$^2$s, hole diameter of $a=30$ nm, and period of $\calP=45$ nm  at $1300$ K, $1700$ K, and $2000$ K.
\label{fig:spectral_radiance_4m} }
\end{figure}

Consequently, we have
\be
\delta\hat{\bE}(\br,\omega)=\omega^2\mu_0\delta\tbp_0(\omega)\sum\limits_j\bG(\br,\bR_j;\omega).
\ee
The dyadic Green function is defined as
\be
\obG(\br,\br';\omega)=\left[\unit+\frac{1}{\bk(\omega)^2}\bnabla\bnabla\right]G(\br,\br';\omega)
\ee
with the scalar Green function given by
\be
G(\br,\br';\omega)=\frac{e^{-i\bk(\omega)\cdot|\br-\br'|}}{4\pi|\br-\br'|},
\ee
and $\bk(\omega)^2=(\omega^2/c^2)[\varepsilon_{xx}(\omega),\varepsilon_{yy}(\omega),\varepsilon_{zz}(\omega)]$.

Then, the fluctuation-dissipation theorem can be recast into the forms
\bea
\left<\delta \tp_\mu(\br_{0||},\omega)\delta \tp_\nu^*(\br_{0||}',\omega')\right> & = & 
\frac{{\sigma'}_{\mu\nu}^{2D}(\bR_i,\omega)}{\omega^2}\Theta(\omega)\delta(\omega-\omega') \nn\\
& & \times\delta(\br_{0||}-\br_{0||}'),
\eea
and thus we obtain
\begin{align}
& \left<\delta\hat{E}_\mu(\br,\omega)\delta\hat{E}_\nu^*(\br',\omega')\right> = 
\omega^4\mu_0^2\sum\limits_{m,m'}\int\limits_{\Omega}d^2r_{0||} G_{\mu m}(\br,\br_{0||};\omega) \nn\\
& \times \int\limits_{\Omega'}d^2r_{0||}'G_{m'\nu}^*(\br',\br_{0||}';\omega')\left<\delta \tp_m(\br_0,\omega)\delta \tp_{m'}^*(\br_0',\omega)\right> \nn\\
& = 
\frac{\omega^2}{c^4\varepsilon_0^2}\sum\limits_{m}\int\limits_{\Omega}d^2r_{0||}G_{\mu m}(\br,\br_{0||};\omega)G_{m'\nu}^*(\br',\br_{0||};\omega') \nn\\
& \times \Theta(\omega){\sigma'}_{mm'}^{2D}(\br_{0||},\omega)\delta(\omega-\omega') \nn\\
& =  
\frac{\omega^2}{c^4\varepsilon_0^2}\sum\limits_{m,j}G_{\mu m}(\br,\bR_j;\omega)G_{m\nu}^*(\br',\bR_j;\omega') \nn\\
& \times \Theta(\omega){\sigma'}_{mm}^{2D}(\bR_j,\omega)\delta(\omega-\omega'),
\end{align}
noting that the dielectric tensor $\varepsilon''(\br,\omega)$ is diagonal.

\begin{figure}[htb]
\begin{centering}
\includegraphics[width=8.5cm]{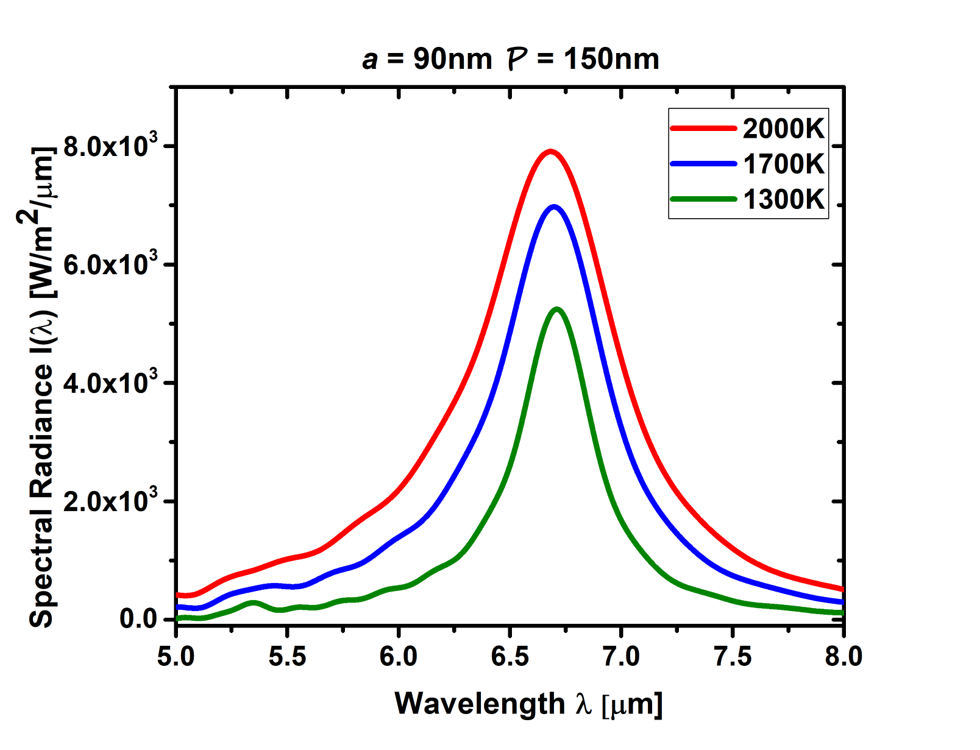}
\end{centering}
\caption{Spectral radiance of NPG including cavity, as shown in in Figs.~\ref{fig:graphene_emitter} and \ref{fig:graphene_emitter_materials},
 as a function of wavelength $\lambda$
with Fermi energy $E_F=-1.0$ eV, mobility $\mu=3000$ V/cm$^2$s, hole diameter of $a=90$ nm, and period of $\calP=150$ nm  at $1300$ K, $1700$ K, and $2000$ K.
\label{fig:spectral_radiance_7m} }
\end{figure}

Since the energy density of the emitted electric field at the point $\br$ is
\be
u(\br,\omega)\delta(\omega-\omega')=\varepsilon_0\sum_{i=x,y,z}\left<\delta\hat{E}_i^*(\br,\omega)\delta\hat{E}_i(\br,\omega')\right>, 
\ee
we can write the spectral radiance as
\bea
I(\br,\omega) & = & \frac{\omega^2}{4\pi c^3\varepsilon_0}\frac{1}{N}\sum\limits_{\mu; m=x,y;j}\left|G_{\mu m}(\br,\bR_j;\omega)\right|^2 \nn\\
& & \times \Theta(\omega){\sigma'}_{mm}^{2D}(\bR_j,\omega) \nn\\
& = &  \frac{\omega^2}{4\pi c^3\varepsilon_0}\Theta(\omega){\sigma'}_{||}^{2D}(\omega) \sum\limits_{\mu,m}\left|G_{\mu m}(\br,\bR_0;\omega)\right|^2,
\label{eq:radiance}
\eea
assuming that the dipole current of the LSP is in the plane of the graphene sheet, i.e. the xy-plane,
and the polarizability is isotropic, ie. ${\sigma'}_{||}^{2D}={\sigma'}_{xx}^{2D}={\sigma'}_{yy}^{2D}$, and the same for all holes.
$N$ is the number of holes.
In order to obtain the spectral radiance in the far field, we need to integrate over the spherical angle.
Using the results from the Supplementary Information, we obtain
\bea
I_\infty(\omega) & = & \frac{\omega^2\Theta(\omega)}{3\pi^2\varepsilon_0 c^3}{\sigma'}_{||}^{2D}(\omega) \nn\\
& = & \frac{\omega^2\Theta(\omega)}{3c^2\pi^2} A_{||}^{2D}(\omega),
\label{eq:radiance0}
\eea
where we used the definition of the absorbance of a 2D material, i.e.
\be
A_{2D}(\omega)=(1/\varepsilon_0 c){\rm Re}\sigma_{2D}(\omega)=(1/\varepsilon_0 c)\sigma_{2D}'(\omega),
\ee
with 2D complex conductivity $\sigma_{2D}(\omega)$.
According to Kirchhoff's law, emittance $\epsilon(\omega)$, absorbance $A(\omega)$, reflectance $R(\omega)$, and transmittance $T(\omega)$ are related by\cite{Lifshitz1980}
\be
\epsilon(\omega)=A(\omega)=1-R(\omega)-T(\omega),
\ee
from which we obtain the grey-body thermal emission formula
\be
I_\infty(\omega) = \frac{\omega^2\Theta(\omega)}{3\pi^2c^2} \epsilon_{||}^{2D}(\omega),
\label{eq:IBB_graphene}
\ee
whose prefactor bears strong similarity to Planck's black body formula in Eq.~(\ref{eq:IBB}).

\begin{figure}[htb]
\begin{centering}
\includegraphics[width=8.5cm]{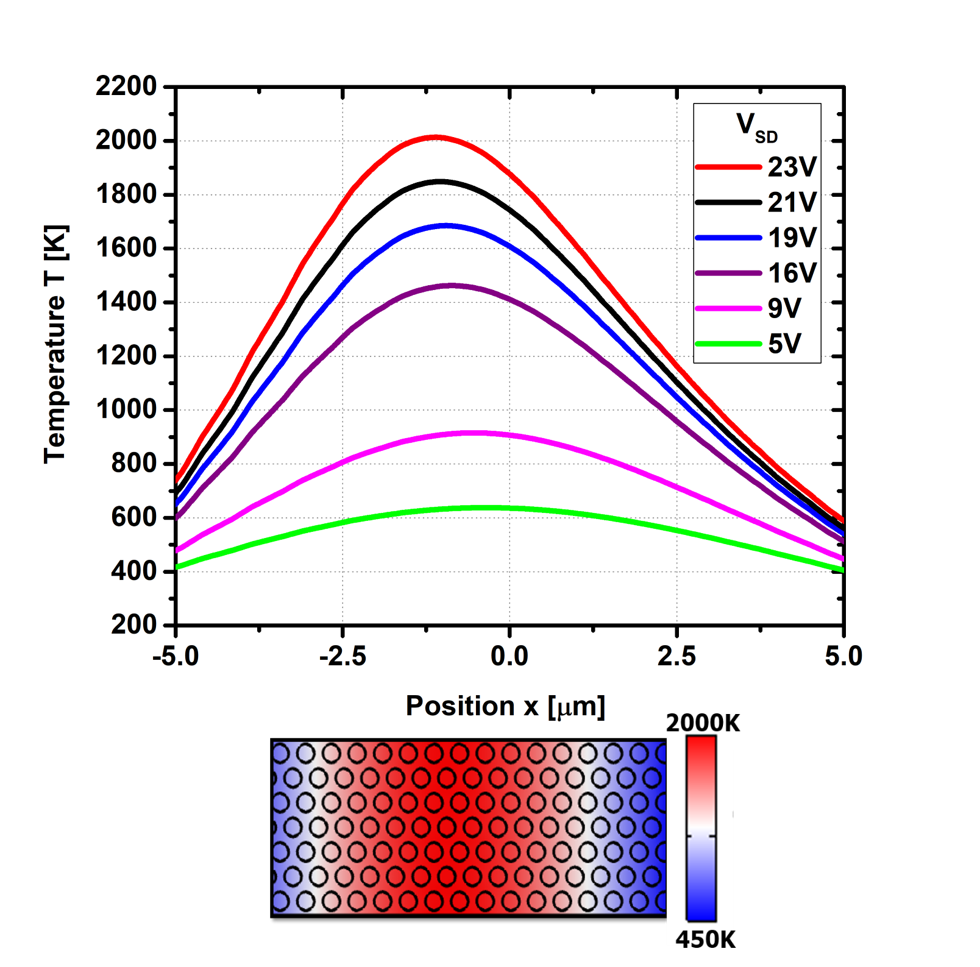}
\end{centering}
\caption{Temperature distribution inside the NPG sheet for various values of the bias voltage $V_{\rm SD}$, calculated by means of COMSOL. As the bias voltage is increased, the maximum of temperature shifts away from the center of the NPG sheet due to the Peltier effect.
\label{fig:Joule_heating_distribution} }
\end{figure}

Using FDTD to calculate the emittance $\epsilon_{||}^{2D}(\omega)$, we evaluted the grey-body thermal emission according to Eq.~(\ref{eq:IBB_graphene})
for the thermal emitter structure based on NPG shown in Figs.~\ref{fig:graphene_emitter} and \ref{fig:graphene_emitter_materials}.
Using COMSOL, we calculated the temperature distribution inside the NPG sheet, as shown in Fig.~\ref{fig:Joule_heating_distribution}, when a bias voltage $V_{\rm SD}$ is applied, which gives rise to Joule heating.
Our results are shown in Figs.~\ref{fig:spectral_radiance_4m}, \ref{fig:spectral_radiance_7m}, and \ref{fig:spectral_radiance_10m} for the temperatures $1300$ K, $1700$ K, and $2000$ K of NPG.
After integrating over the wavelength under the curves, we obtain the following thermal emission power per area:
\begin{center}
\begin{tabular}{|c|c|}\hline
Resonance wavelength & Power per area \\ \hline\hline
4 $\mu$m & 11,221 W/m$^2$ \\ \hline
7 $\mu$m & 9820 W/m$^2$ \\ \hline
10 $\mu$m & 6356 W/m$^2$ \\ \hline
\end{tabular}
\end{center}

\begin{figure}[htb]
\begin{centering}
\includegraphics[width=8.5cm]{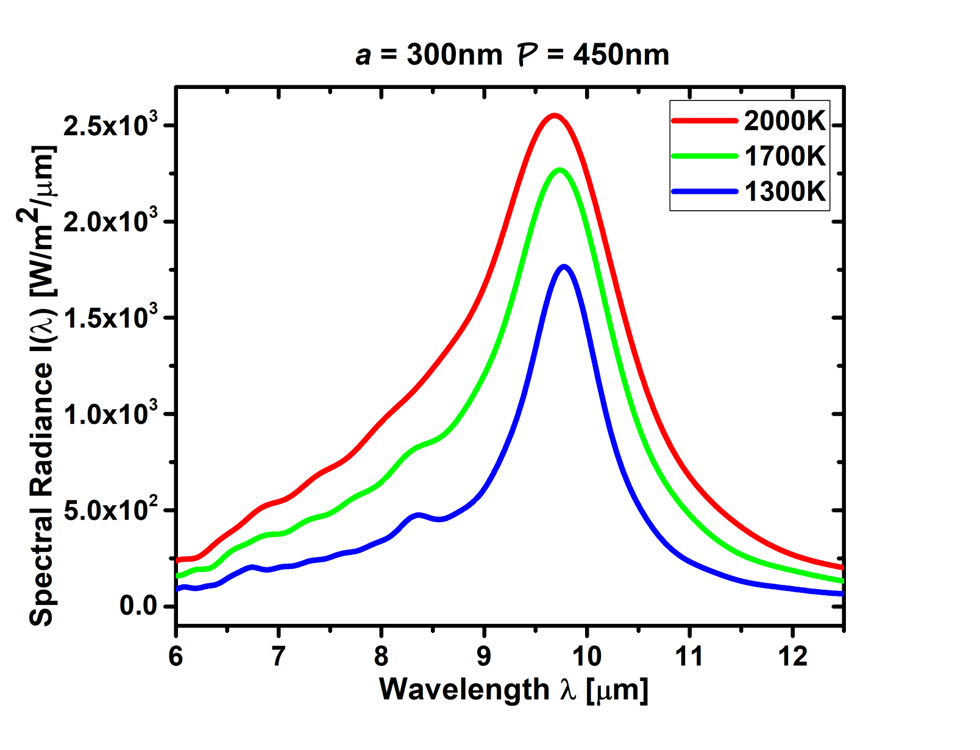}
\end{centering}
\caption{Spectral radiance of NPG including cavity, as shown in in Figs.~\ref{fig:graphene_emitter} and \ref{fig:graphene_emitter_materials},
 as a function of wavelength $\lambda$
with Fermi energy $E_F=-1.0$ eV, mobility $\mu=3000$ V/cm$^2$s, hole diameter of $a=300$ nm, and period of $\calP=450$ nm  at $1300$ K, $1700$ K, and $2000$ K.
\label{fig:spectral_radiance_10m} }
\end{figure}

Let us consider the dependence of the thermal emission of NPG on the angle $\theta$.
Integrating over $r^2\varphi$ we obtain
\be
I(\theta,\omega)=\frac{\omega^2}{4\pi c^2}\Theta(\omega)\frac{11+\cos(2\theta)}{16\pi}\epsilon_{||}^{2D}(\omega),
\ee
which is a clear deviation from a Lambert radiator.
The pattern of the thermal radiation can be determined by
\bea
\hat{I}(\theta) & = & \frac{\int_0^{2\pi}I(\br,\omega)r^2 d\varphi}{\int_0^{2\pi}\int_0^{\pi}I(\br,\omega)r^2\sin\theta d\theta d\varphi} \nn\\
& = & \frac{3}{64}\left[11+\cos(2\theta)\right],
\eea
which is shown in Fig.~\ref{fig:intensity_angular_distribution}.
Interestingly, since we assumed that thermal emission is completely incoherent [see Eq.~(\ref{eq:radiance})] the thermal emission from NPG is only weakly dependent on the emission angle $\theta$, which can be clearly seen in  Fig.~\ref{fig:intensity_angular_distribution}.

\begin{figure}[htb]
\begin{centering}
\includegraphics[width=7.5cm]{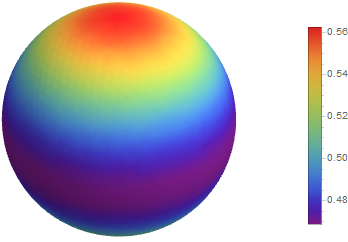}
\end{centering}
\caption{Spherical density plot of the normalized angular intensity distribution $\hat{I}(\theta)$ of the thermal emission from NPG in the case of incoherent photons.
\label{fig:intensity_angular_distribution} }
\end{figure}

However, the assumption that thermal emission of radiation is incoherent is not always true. Since Kirchhoff's law is valid, thermal sources can be coherent.\cite{Greffet2017}
After theoretical calculations predicted that long-range coherence may exist for thermal emission in the case of resonant surface waves, either plasmonic or phononic in nature,\cite{Carminati1999,Henkel2000} experiments showed that a periodic microstructure in the polar material SiC exhibits coherence over many wavelengths and radiates in well-defined and controlled directions.\cite{Greffet2002}
Here we show that the coherence length of a graphene sheet patterned with circular holes can be as large as 150 $\mu$m due to the plasmonic wave in the graphene sheet, thereby paving the way for the creation of phased arrays made of nanoantennas represented by the holes in NPG.

The coherence of thermal emission can be best understood by means of a nonlocal response function.\cite{Henkel2006}
First, we choose the nonlocal hydrodynamic response function in Eq.~(\ref{eq:sigma_HD}).
Using the 2D version of the fluctuation-dissipation theorem in Eq.~(\ref{eq:fluctuation-disspation_2D}), we obtain the nonlocal fluctuation-dissipation theorem in the hydrodynamic approximation,
\begin{align}
\left<\delta\hat{J}_\mu(\br_{||},\omega)\delta\hat{J}_\nu(\br_{||}',\omega')\right> = \sigma_{\mu\nu}^{\rm HD}(\Delta\br_{||},\omega)\Theta(\omega)\delta(\omega-\omega') &
 \nn\\
=\frac{1}{D}\int\limits_0^\infty dk_{||} \frac{\sigma_{\rm intra}(\omega)e^{-ik_{||}\Delta\br_{||}}}{1-\eta^2\frac{k_{||}^2}{\omega^2}}
\Theta(\omega)\delta(\omega-\omega') & \nn\\
 =\sigma_{\rm intra}(\omega) \frac{\omega\sqrt{\pi/2}}{D\eta}\sin\left(\frac{\omega\Delta\br_{||}}{\eta}\right) \Theta(\omega)\delta(\omega-\omega'), &
\label{eq:fluctuation-disspation_2D_hydro}
\end{align}
where $\Delta\br_{||}=\br_{||}-\br_{||}'$ and $\eta^2=\beta^2+D^2\omega(\gamma+i\omega)$.
This result suggests that the coherence length is given approximately by $D$, which according to Ref.~\onlinecite{Bandurin2016} would be $D\approx 0.4$ $\mu$m.
However, the resulting broadening of the LSP resonance peaks would be very large and therefore in complete contradiction to the experimental measurements of the LSP resonance peaks in Refs.~\onlinecite{Safaei2017,SafaeiACS,Safaei2019}.
Thus, we conclude that the hydrodynamic diffusion length must be frequency-dependent with $D(\nu=0)=0.4$ $\mu$m.
Using the Fermi velocity of $v_F=10^6$ m/s and a frequency of $\nu=30$ THz, the average oscillation distance is about $L=v_F\nu^{-1}=0.033$ $\mu$m, which is much smaller than $D(\nu=0)$ in graphene. Thus we can approximate $D(\nu=30$ THz$)=0$.
We conjecture that there is a crossover for $D$ into the hydrdynamic regime when the frequency is reduced below around $\nu_0=1$ to $3$ THz, below which the hydrodynamic effect leads to a strong broadening of the LSP peaks for NPG. Consequently, the viscosity of graphene should also be frequency-dependent and a crossover for the viscosity should happen at about the same frequency $\nu_0$. We plan to elaborate this conjecture in future work. Future experiments could corroborate our conjecture by measuring the absorbance or emittance as a function of wavelength for varying scale of patterning of the graphene sheet.

\begin{figure}[htb]
\begin{centering}
\includegraphics[width=8.5cm]{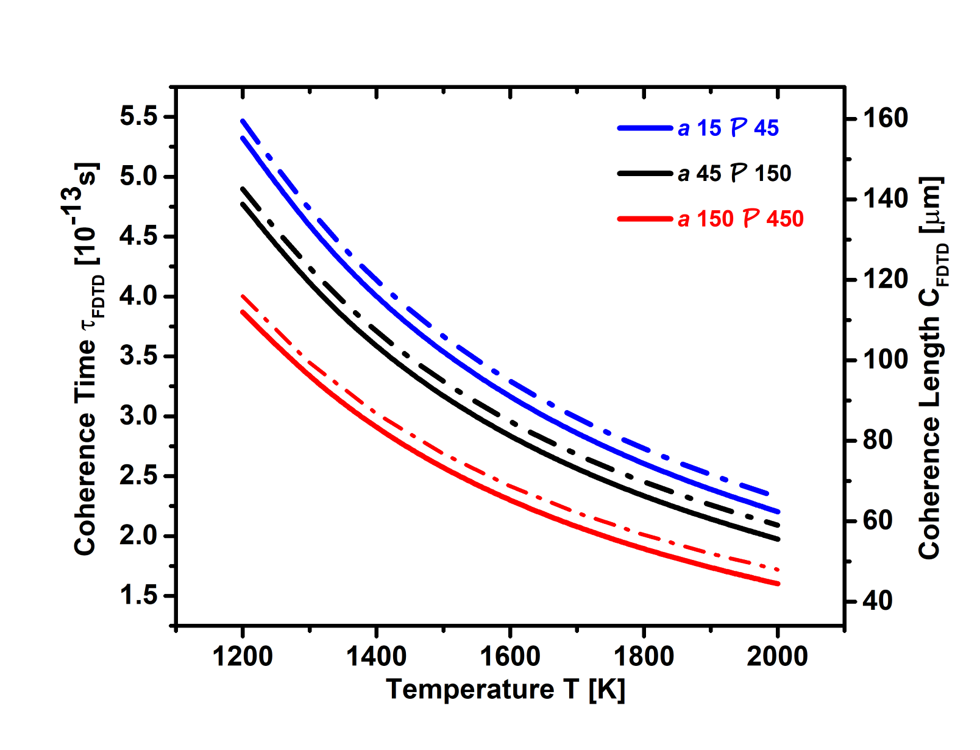}
\end{centering}
\caption{Coherence length $C_{\rm FDTD}$ and coherence time $\tau_{\rm FDTD}$ of emitted photons, extracted from the full-width half-maximum (FWHM) of the spectral radiances shown in Figs.~\ref{fig:spectral_radiance_4m}, \ref{fig:spectral_radiance_7m}, and \ref{fig:spectral_radiance_10m}. 
\label{fig:coherence} }
\end{figure}

Next, let us consider the coherence of thermal emission by means of the nonlocal optical conductivity in the RPA approximation.
Using the general formula
\be
\sigma(q,\omega)=\frac{ie^2\omega}{q^2}\chi^0(q,\omega),
\ee
with
\be
\chi^0(q,\omega)\approx \frac{\epsilon_F q^2}{\pi\hbar^2\omega(\omega+i\tau^{-1})}
\ee
in the low-temperature and low-frequency approximation, one obtains Eq.~(\ref{eq:sigma_intra}).
Now, let us use the full polarization in RPA approximation including only the Coulomb interaction,
\be
\chi^{\rm RPA}(q,\omega)=\frac{\chi^0(q,\omega)}{1- v_c(q){\chi^0}(q,\omega ) },
\ee
from which we obtain
\bea
\sigma^{\rm RPA}(q,\omega) & = & \frac{ie^2\omega}{q^2}\chi(q,\omega) \nn\\
& = &  \frac{ie^2\omega\epsilon_F}{\pi\hbar^2\omega(\omega+i\tau^{-1})-\frac{e^2\epsilon_F}{2\epsilon_0} q},
\eea
which introduces the nonlocal response via the Coulomb interaction in the denominator. After taking the Fourier transform, we obtain the nonlocal fluctuation-dissipation theorem in RPA approximation,
\begin{align}
\left<\delta\hat{J}_\mu(\br_{||},\omega)\delta\hat{J}_\nu(\br_{||}',\omega')\right> = \sigma_{\mu\nu}^{\rm RPA}(\Delta\br_{||},\omega)\Theta(\omega)\delta(\omega-\omega') & \nn\\
 =\frac{\sqrt{2\pi}\epsilon_0\omega}{C_{RPA}} e^{iK_{RPA}\Delta\br_{||}-\frac{\Delta\br_{||}}{C_{RPA}}} \Theta(\omega)\delta(\omega-\omega'), &
\label{eq:fluctuation-disspation_2D_RPA}
\end{align}
where the coherence length in RPA approximation is
\be
C_{RPA}=\frac{e^2|\epsilon_F|}{2\pi\hbar^2\epsilon_0\gamma\omega},
\ee
and the coherence wavenumber is given by 
\be
K_{RPA}=\frac{2\pi\hbar^2\epsilon_0\omega^2}{e^2|\epsilon_F|}.
\ee
For simplicity, we switch now to a square lattice of holes.
In the case of the LSP resonance for a square lattice of holes at $\lambda=10$ $\mu$m, corresponding to $\nu=30$ THz, $\epsilon_F=-1.0$ eV, $\omega=2\pi\nu$, and $\gamma=ev_F^2/(\mu E_F)=0.3$ THz for $\mu=3000$ cm$^2$V$^{-1}$s$^{-1}$, which results in a coherence length of $C_{RPA}=3$ $\mu$m.
This result is in reasonable agreement with the full width at half maximum (FWHM) values of the widths of the LSP resonance peaks in Refs.~\onlinecite{Safaei2017,SafaeiACS,Safaei2019}.
This coherence length would allow to preserve coherence for a linear array of period $\calP=300$ nm and $C_{RPA}/\calP=10$ holes.
In order to show the coherence length that can be achieved with graphene, we can consider a suspended graphene sheet with a mobility of $\mu=15000$ cm$^2$V$^{-1}$s$^{-1}$.
Then the coherence length increases to a value of $C_{RPA}=13$ $\mu$m, which would allow for coherence over a linear array with $C_{RPA}/\calP=43$ holes.

\begin{figure}[htb]
\begin{centering}
\includegraphics[width=8.5cm]{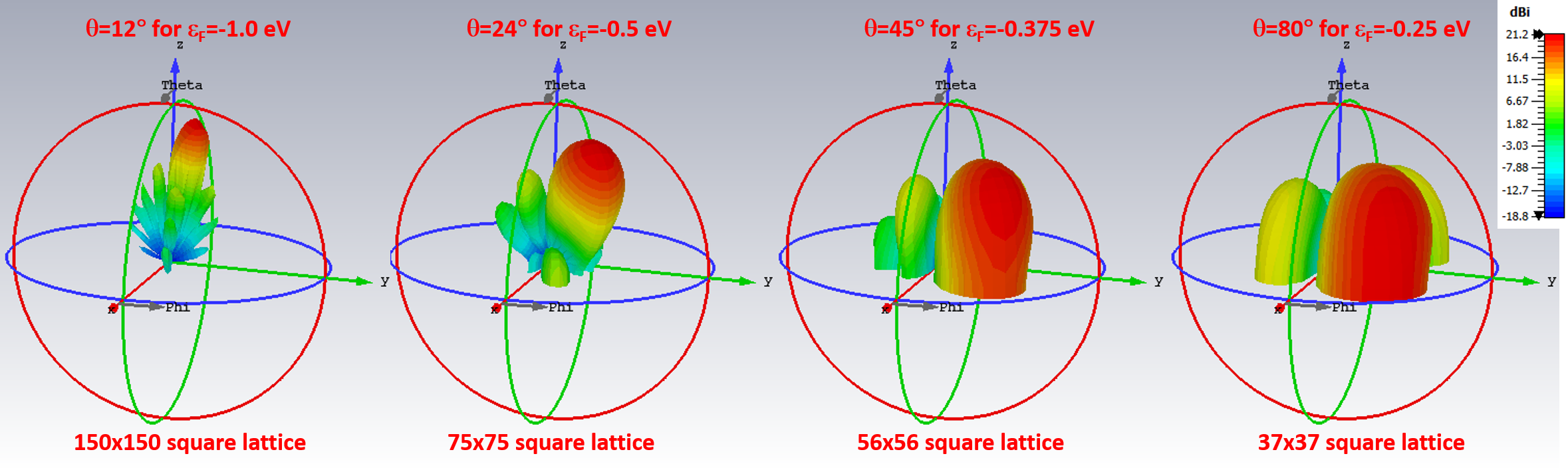}
\end{centering}
\caption{Directivity of the thermal emission from NPG where the holes act as nanoantennas in a phased array. This emission pattern for $\epsilon_F=-1.0$ eV can be used for surface-emitting mid-IR sources. In the case of a 150x150, 75x75, 56x56, 37x37 square lattice of holes (size of lattice matches coherence length) with period $\calP=45$ nm and hole diameter of $30$ nm, introducing a relative phase of $2.43^\circ$, $4.86^\circ$, $7.28^\circ$, $9.71^\circ$ between the nanoantennas allows for beamsteering in the range between $\theta=12^\circ$ and $\theta=80^\circ$ by tuning the Fermi energy in the range between $\epsilon_F=-1.0$ eV and $\epsilon_F=-0.25$ eV.
\label{fig:phase_array_beamsteering} }
\end{figure}

In the case of the LSP resonance for a square lattice of holes at $\lambda=5$ $\mu$m, corresponding to $\nu=60$ THz, $\epsilon_F=-1.0$ eV, $\omega=2\pi\nu$, and $\gamma=ev_F^2/(\mu E_F)=0.3$ THz for $\mu=3000$ cm$^2$V$^{-1}$s$^{-1}$ , which results in a coherence length of $C_{RPA}=1.5$ $\mu$m.
Considering again a suspended graphene sheet, the coherence length can be increased to $C_{RPA}=6.7$ $\mu$m.
Since the period in this case is $\calP=45$ nm, the coherence for $\mu=3000$ cm$^2$V$^{-1}$s$^{-1}$ and $\mu=15000$ cm$^2$V$^{-1}$s$^{-1}$  can be preserved for a linear array of $C_{RPA}/\calP=33$  and 148 holes, respectively.

The coherence length and time of thermally emitted photons is larger because the photons travel mostly in vacuum. Taking advantage of the Wiener-Kinchine theorem,\cite{Greffet2017} we can extract the
coherence length $C_{\rm FDTD}$ and coherence time $\tau_{\rm FDTD}$ of thermally emitted photons by means of the full-width half-maximum (FWHM) of the spectral radiances shown in Figs.~\ref{fig:spectral_radiance_4m}, \ref{fig:spectral_radiance_7m}, and \ref{fig:spectral_radiance_10m}. 
Our results are shown in Fig.~\ref{fig:coherence}. The coherence length of the thermally emitted photons can reach up to $C_{\rm FDTD}=150$ $\mu$m at a resonance wavelength of $\lambda=4$ $\mu$m. This means that the coherence length of the thermally emitted photons is about 37 times larger than the wavelength.

Thus, the latter large coherence length allows for the coherent control of a 150x150 square array of holes with period $\calP=45$ nm, individually acting as nanoantennas, that can be used to create a phased array of nanoantennas. 
One of the intriguing properties of a phased array is that it allows to control the directivity of the emission of photons, which is currently being implemented for large 5G antennas in the 3 to 30 GHz range. The beamsteering capability of our NPG sheet is shown in Fig.~\ref{fig:phase_array_beamsteering}.
In contrast, our proposed phased array based on NPG can operate in the 10 to 100 THz range.

The temporal control of the individual phases of the holes requires an extraordinary fast switching time of around 1 ps, which is not feasible with current electronics.
However, the nonlocal response function reveals a spatial phase shift determined by the coherence wavenumber $K_{RPA}$, which is independent of the mobility of graphene.
In the case of the LSP resonance at $\lambda=4$ $\mu$m, we obtain $\lambda_{RPA}=2\pi/K_{RPA}=6$ $\mu$m, resulting in a minimum phase shift of $2\pi\calP/\lambda_{RPA}=0.042=2.4^{\circ}$ between neighboring holes, which can be increased to a phase shift of $9.7^{\circ}$ by decreasing the Fermi energy to $E_F=-0.25$ eV.
Thus, the phase shift between neighboring holes can be tuned arbitrarily between $2.4^{\circ}$ and $9.7^{\circ}$  by varying the Fermi energy  between $\epsilon_F=-1.0$ eV and $\epsilon_F=-0.25$ eV.
Fig.~\ref{fig:phase_array_beamsteering} shows the capability of beamsteering for our proposed structure by means of directional thermal emission, which is tunable by means of the gate voltage applied to the NPG sheet.

Due to the full control of directivity with angle of emission between $\theta=12^\circ$ and $\theta=80^\circ$ by tuning the Fermi energy in the range between $\epsilon_F=-1.0$ eV and $\epsilon_F=-0.25$ eV, thereby achieving beamsteering by means of the gate voltage, our proposed mid-IR light source based on NPG can be used not only in a vertical setup for surface emission, but also in a horizontal setup for edge emission, which is essential for nanophotonics applications.

In conclusion, we have demonstrated in our theoretical study that NPG can be used to develop a plasmonically enhanced mid-IR light source with spectrally tunable selective thermal emission.
Most importantly, the LSPs along with an optical cavity increase substantially the emittance of graphene from about 2\% for pristine graphene to 80\% for NPG, thereby outperforming state-of-the-art graphene light sources working in the visible and NIR by at least a factor of 100.
Combining our proposed mid-IR light source based on patterned graphene with our demonstrated mid-IR detector based on NPG\cite{Safaei2019},
we are going to develop a mid-IR spectroscopy and detection platform based on patterned graphene 
that will be able to detect a variety of molecules that have mid-IR vibrational resonances, such as
CO, CO$_2$, NO, NO$_2$, CH$_4$, TNT, H$_2$O$_2$, acetone, TATP, Sarin, VX, etc.
In particular, a recent study showed that it is possible to detect the hepatitis B and C viruses
label-free at a wavelength of around 6 $\mu$m.\cite{Roy2019}
Therefore, we will make great effort to demonstrate that our platform will be able to detect with high sensitivity and selectivity the COVID-19 virus and other viruses that pose a threat to humanity.

\begin{acknowledgments}
\textcolor{black}{We acknowledge support from NSF CISE-1514089.
We thank Gernot Pomrenke, Alireza Safaei, and Sayan Chandra for useful discussions.}
\end{acknowledgments}

\section{Supplementary Information}

\subsection{Spectrally Selective Thermal Emission}

\begin{figure}[htb]
\begin{centering}
\includegraphics[width=8.5cm]{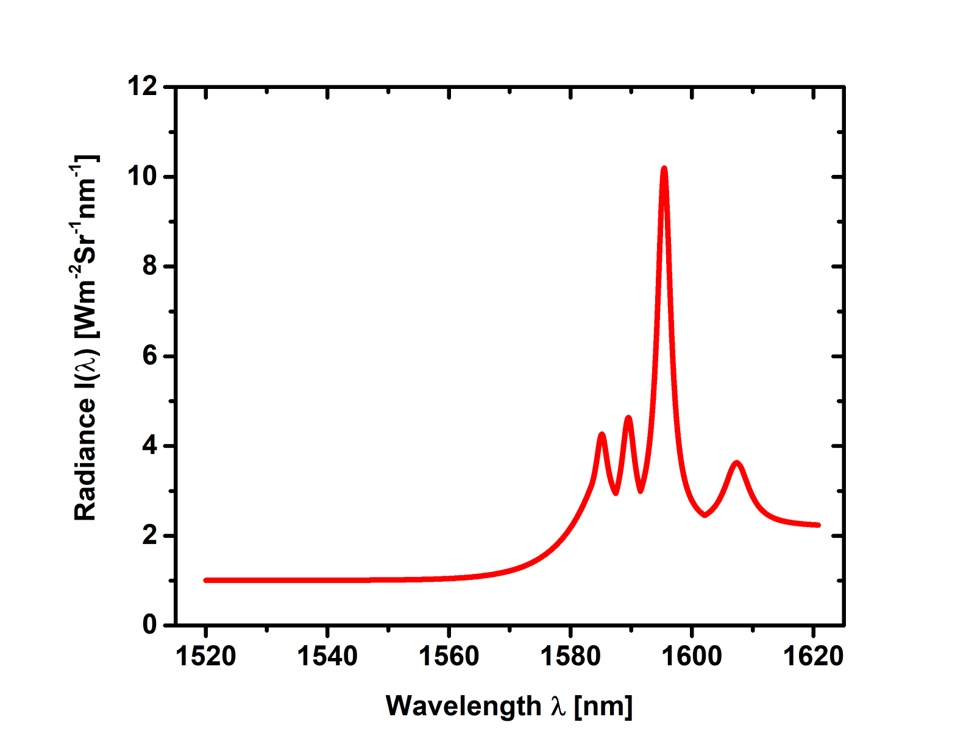}
\end{centering}
\caption{Theoretical fit to spectral radiance presented in Ref.~\onlinecite{Shiue2019}. Shiue et al. used a photonic crystal structure to filter the thermal emission from pristine graphene with an emittance of around $A=0.07$. Integrating the spectral radiance under the curve gives a value of about $P/A=100$ W/m$^2$, which is about 100 times weaker than our proposed thermal radiation source based on NPG.
\label{fig:spectral_radiance_Shiue} }
\end{figure}

Kirchhoff's law of thermal radiation states that emittance $\epsilon$ is equal to absorbance $A$, i.e.
\be
\epsilon(\omega,\theta,\phi,T)=A(\omega,\theta,\phi,T).
\ee
In the case of a black body $\epsilon(\omega,\theta,\phi,T)=A(\omega,\theta,\phi,T)=1$.
Pristine graphene has a very small absorbance of only $A=0.023$ and is a nearly transparent body.
Shiue et al. used a photonic crystal structure to filter the thermal emission from pristine graphene with an emittance of around $A=0.07$.\cite{Shiue2019}
Their spectral radiance is shown in Fig.~\ref{fig:spectral_radiance_Shiue} and exhibits peaks at around $\lambda=1.55$ $\mu$m at a temperature of $T=2000$ K.
After integrating the spectral radiance under the curve, one obtains a emission power per area of about $P/A=100$ W/m$^2$, which is about 100 times weaker than our proposed thermal radiation source based on NPG at $T=2000$ K.
Our proposed thermal mid-IR source features an emission power per area of about $P/A=10^4$ W/m$^2$ at $T=2000$ K. In addition, our proposed thermal mid-IR source features frequency-tunability and beamsteering by means of a gate voltage applied to the NPG sheet.

\begin{figure}[htb]
\begin{centering}
\includegraphics[width=8.5cm]{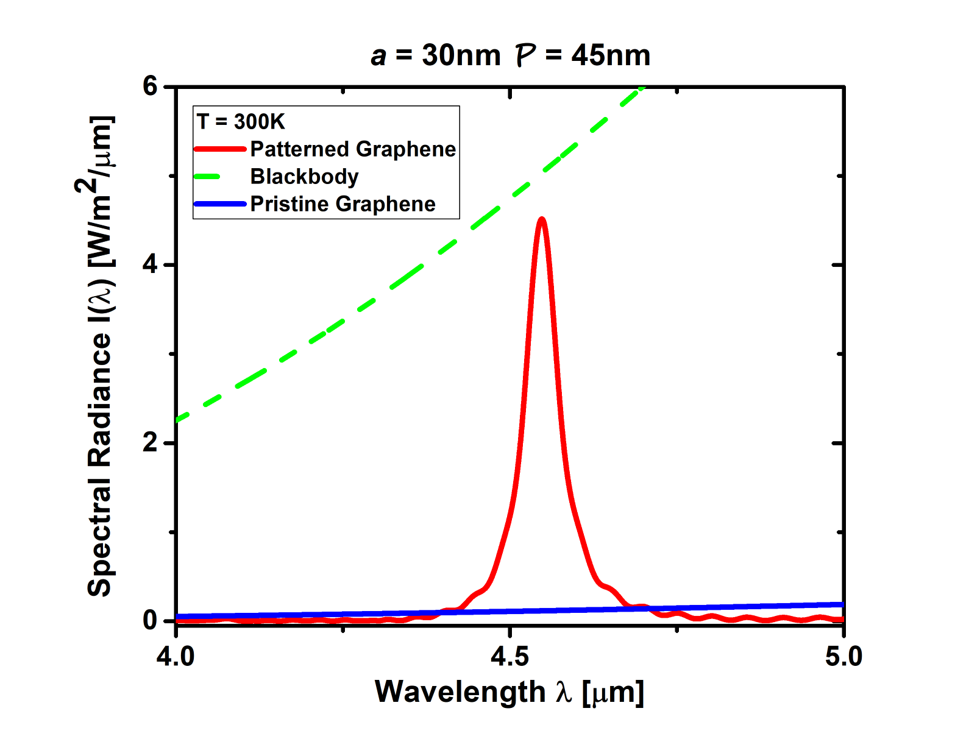}
\end{centering}
\caption{The NPG sheet allows for spectrally selective thermal emission at around $\lambda=4.5$ $\mu$m for a period of $\calP=45$ nm and a hole diameter of $a=30$ nm.
\label{fig:spectral_radiance_4m_300K} }
\end{figure}

Using FDTD to calculate the emittance $\epsilon_{||}^{2D}(\omega)$, we evaluted the grey-body thermal emission according to Eq.~(\ref{eq:IBB_graphene})
for the thermal emitter structure based on NPG shown in Figs.~\ref{fig:graphene_emitter} and \ref{fig:graphene_emitter_materials}.
Our results for the temperature $T=300$ K  of NPG are shown in Figs.~\ref{fig:spectral_radiance_4m_300K}, \ref{fig:spectral_radiance_7m_300K}, and \ref{fig:spectral_radiance_10m_300K}.
In these figures we compare our results for NPG with the results for pristine graphene and black body radiation.

\begin{figure}[htb]
\begin{centering}
\includegraphics[width=8.5cm]{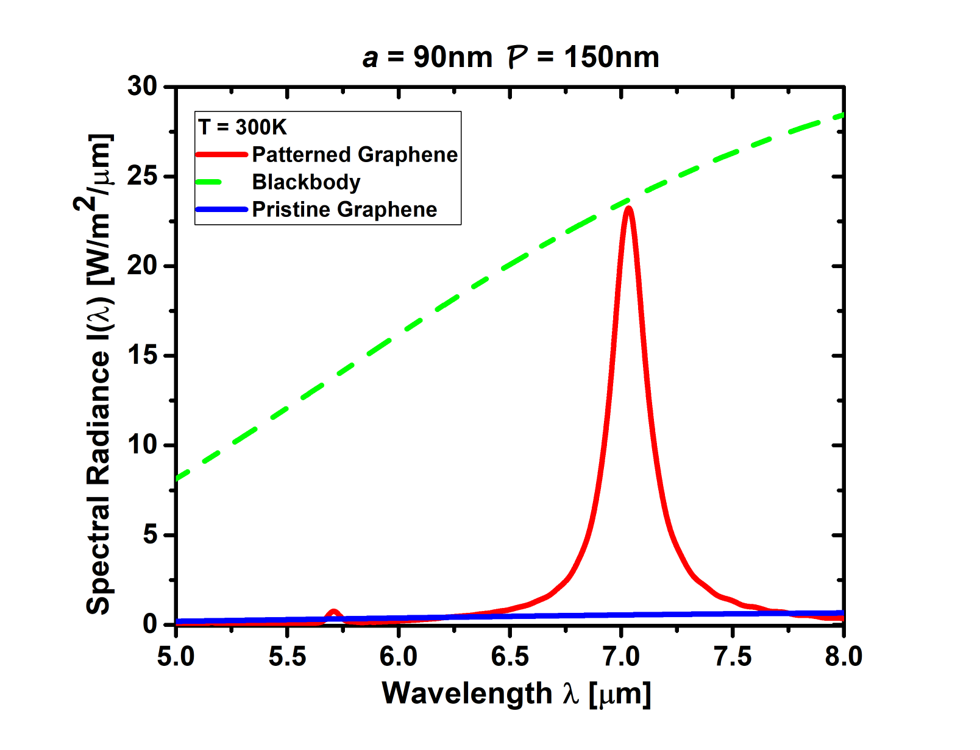}
\end{centering}
\caption{The NPG sheet allows for spectrally selective thermal emission at around $\lambda=7$ $\mu$m for a period of $\calP=150$ nm and a hole diameter of $a=90$ nm.
\label{fig:spectral_radiance_7m_300K} }
\end{figure}

\subsection{Ellipsoidal Coordinates}
For determining the EM properties of an infinitesimally thin conducting elliptical disk of radius $R$ or an infinitesimally thin conducting plane with a elliptical hole, including coated structures,
it is most convenient to perform the analytical calculations in the ellipsoidal coordinate system ($\xi$, $\eta$, $\zeta$),\cite{AbramowitzStegun,Li2002,Landau1984,Bohren1998} which is
related to the Cartesian coordinate system through the implicit equation
\be
\frac{x^2}{a^2+u}+\frac{y^2}{b^2+u}+\frac{z^2}{c^2+u}=1
\label{eq:ellipsoid}
\ee
for $a>b>c$. The cubic roots $\xi$, $\eta$, and $\zeta$ are all real in the ranges
\be
-a^2\le\zeta\le -b^2,\; -b^2\le\eta\le -c^2,\; -c^2\le\xi <\infty,
\ee
which are the ellipsoidal coordinates of a point $(x,y,z)$.
The surfaces of contant $\xi$, $\eta$, and $\zeta$ are ellipsoids, hyperboloids of one sheet, and hyperboloids of two sheets, respectively,
all confocal with the ellipsoid defined by
\be
\frac{x^2}{a^2}+\frac{y^2}{b^2}+\frac{z^2}{c^2}=1.
\ee
Each point $(x,y,z)$ in space is determined by the intersection of three surfaces, one from each of the three families, and the three surfaces are orthogonal to each other.
The transformation between the two coordinate systems is given by the solutions of Eq.~(\ref{eq:ellipsoid}), i.e.
\bea
x & = & \pm\sqrt{\frac{(\xi+a^2)(\eta+a^2)(\zeta+a^2)}{(b^2-a^2)(c^2-a^2)}}, \\
y & = & \pm\sqrt{\frac{(\xi+b^2)(\eta+b^2)(\zeta+b^2)}{(c^2-b^2)(a^2-b^2)}}, \\
z & = & \pm\sqrt{\frac{(\xi+c^2)(\eta+c^2)(\zeta+c^2)}{(a^2-c^2)(b^2-c^2)}},
\eea 
defining 8 equivalent octants.
The length elements in ellipsoidal coordinates read
\bea
dl^2 & = & h_1^2d\xi^2+h_2^2 d\eta^2+h_3^2 d\zeta^2, \\
h_1 & = & \sqrt{\frac{(\xi-\eta)(\xi-\zeta)}{2R_\xi}}, \\
h_2 & = & \sqrt{\frac{(\eta-\zeta)(\xi-\zeta)}{2R_\eta}}, \\
h_3 & = & \sqrt{\frac{(\zeta-\xi)(\zeta-\eta)}{2R_\zeta}}, \\
R_u^2 & = & (u+a^2)(u+b^2)(u+c^2), \; u=\xi,\eta,\zeta.
\eea
For the transformation from cartesian to ellipsoidal coordinates, one can use the following system of equations:
\bea
\hbxi & = & \frac{\frac{\p x}{\p\xi}\hbx + \frac{\p y}{\p\xi}\hby + \frac{\p z}{\p\xi}\hbz}{\sqrt{\left(\frac{\p x}{\p\xi}\right)^2+\left(\frac{\p y}{\p\xi}\right)^2+\left(\frac{\p z}{\p\xi}\right)^2}} , \\
\hbeta & = & \frac{\frac{\p x}{\p\eta}\hbx + \frac{\p y}{\p\eta}\hby + \frac{\p z}{\p\eta}\hbz}{\sqrt{\left(\frac{\p x}{\p\eta}\right)^2+\left(\frac{\p y}{\p\eta}\right)^2+\left(\frac{\p z}{\p\eta}\right)^2}} , \\
\hbzeta & = & \frac{\frac{\p x}{\p\zeta}\hbx + \frac{\p y}{\p\zeta}\hby + \frac{\p z}{\p\zeta}\hbz}{\sqrt{\left(\frac{\p x}{\p\zeta}\right)^2+\left(\frac{\p y}{\p\zeta}\right)^2+\left(\frac{\p z}{\p\zeta}\right)^2}}, 
\eea 
whose elements $J_{ij}$ define the Jacobian matrix. The derivatives are explicitly:
\bea
\frac{\p x}{\p\xi} & = & \frac{1}{2}\sqrt{\frac{(a^2+\eta)(a^2+\zeta)}{(a^2+\xi)(a^2-b^2)(a^2-c^2)}} , \\
\frac{\p x}{\p\eta} & = & \frac{1}{2}\sqrt{\frac{(a^2+\xi)(a^2+\zeta)}{(a^2+\eta)(a^2-b^2)(a^2-c^2)}} ,  \\
\frac{\p x}{\p\zeta} & = & \frac{1}{2}\sqrt{\frac{(a^2+\xi)(a^2+\eta)}{(a^2+\zeta)(a^2-b^2)(a^2-c^2)}} ,  \\
\frac{\p y}{\p\xi} & = & \frac{1}{2}\sqrt{\frac{(b^2+\eta)(b^2+\zeta)}{(b^2+\xi)(b^2-a^2)(b^2-c^2)}} , \\
\frac{\p y}{\p\eta} & = & \frac{1}{2}\sqrt{\frac{(b^2+\xi)(b^2+\zeta)}{(b^2+\eta)(b^2-a^2)(b^2-c^2)}} ,  \\
\frac{\p y}{\p\zeta} & = & \frac{1}{2}\sqrt{\frac{(b^2+\xi)(b^2+\eta)}{(b^2+\zeta)(b^2-a^2)(b^2-c^2)}} ,  \\
\frac{\p z}{\p\xi} & = & \frac{1}{2}\sqrt{\frac{(c^2+\eta)(c^2+\zeta)}{(c^2+\xi)(c^2-a^2)(c^2-b^2)}} , \\
\frac{\p z}{\p\eta} & = & \frac{1}{2}\sqrt{\frac{(c^2+\xi)(c^2+\zeta)}{(c^2+\eta)(c^2-a^2)(c^2-b^2)}} ,  \\
\frac{\p z}{\p\zeta} & = & \frac{1}{2}\sqrt{\frac{(c^2+\xi)(c^2+\eta)}{(c^2+\zeta)(c^2-a^2)(c^2-b^2)}} .
\eea

\begin{figure}[htb]
\begin{centering}
\includegraphics[width=8.5cm]{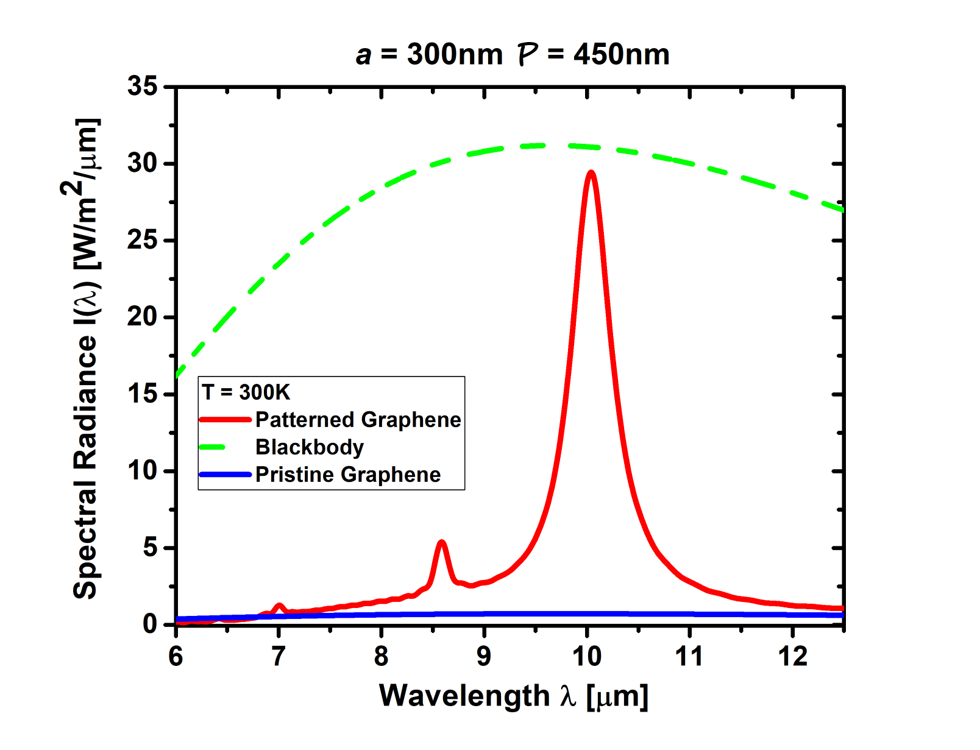}
\end{centering}
\caption{The NPG sheet allows for spectrally selective thermal emission at around $\lambda=10$ $\mu$m for a period of $\calP=450$ nm and a hole diameter of $a=300$ nm.
\label{fig:spectral_radiance_10m_300K} }
\end{figure}

The coordinate $\eta$ is constant on the surfaces of oblate spheroids defined by
\be
\frac{x^2+y^2}{(R\cosh\eta)^2}+\frac{z^2}{(R\sinh\eta)^2}=1
\ee
The surface associated with the limit $\eta\rightarrow 0$ is an infinitesimally thin circular disk of radius $R$.
In contrast, the surface in the limit $\eta\gg 1$ is a sphere of radius $r=R\cosh\eta\approx R\sinh\eta$.
Thus, the Laplace equation in ellipsoidal coordinates reads
\begin{align}
\Delta\Phi = \frac{4}{(\xi-\eta)(\zeta-\xi)(\eta-\zeta)}
\left[
(\eta-\zeta)R_\xi\frac{\p}{\p\xi}\left(R_\xi\frac{\p\Phi}{\p\xi}\right)
\right.\nn\\
+\left.
(\zeta-\xi)R_\eta\frac{\p}{\p\eta}\left(R_\eta\frac{\p\Phi}{\p\eta}\right)
+(\xi-\eta)R_\zeta\frac{\p}{\p\zeta}\left(R_\zeta\frac{\p\Phi}{\p\zeta}\right)
\right]=0.
\end{align}

\subsection{Charged Conducting Ellipsoid}
The surface of the conducting ellipsoid is defined by $\xi=0$.
Thus, the electric field potential $\Phi(\xi)$ is a function of $\xi$ only, thereby defining the equipotential surfaces by confocal ellipsoids.
Laplace's equation is then simplified to
\be
\frac{d}{d\xi}\left(R_\xi\frac{d\Phi}{d\xi}\right)=0.
\ee
The solution outside the ellipsoid is
\be
\Phi_{\rm out}(\xi)=A\int\limits_\xi^\infty\frac{d\xi'}{R_{\xi'}}.
\ee
From the asymptotic approximation $\xi\approx r^2$ for large distances $r\rightarrow\infty$, i.e. $\xi\rightarrow\infty$,
we identify $R_\xi\approx\xi^{3/2}$ and thus
\be
\Phi_{\rm out}(\xi\rightarrow\infty)\approx\frac{2A}{\sqrt{\xi}}=\frac{2A}{r}.
\ee
using the boundary condition $\lim_{\xi\rightarrow\infty}\Phi(\xi)=0$.
Since the Coulomb field should be $\Phi(\xi\rightarrow\infty)\approx e/r$ at large distances from the ellipsoid, $2A=e$ and
\be
\Phi_{\rm out}(\xi)=\frac{e}{2}\int\limits_\xi^\infty\frac{d\xi'}{R_{\xi'}}
\ee
is obtained, corresponding to the far-field of a monopole charge.

The solution inside the ellipsoid is
\be
\Phi_{\rm in}(\xi)=B\int\limits_{-c^2}^\xi\frac{d\xi'}{R_{\xi'}}.
\ee
Using the asymptotic approximation $R_{\xi\rightarrow -c^2}\propto\sqrt{\xi+c^2}$
we obtain
\be
\Phi_{\rm in}(\xi\rightarrow -c^2)\approx B\sqrt{\xi+c^2}.
\ee
This solution satisfies the boundary condition $\lim_{\xi\rightarrow -c^2}\Phi(\xi)=0$.
The constant $B$ can be found from the boundary condition $\Phi(\xi=0)=V$, where V is the potential on the surface of the charged ellipsoid.
Thus, $B=V/c$ and
\be
\Phi_{\rm in}(\xi)=\frac{V}{c}\sqrt{\xi+c^2}.
\ee

\subsection{Dipole Moment of Conducting Ellipsoid induced by an External Electric Field in $z$-direction}
Following Ref.~\onlinecite{Bohren1998}, let us consider the case when the external electric field is parallel to one of the major axes of the ellipsoid.
For the external potential let us choose
\be
\Phi_0=-E_0 z=-E_0 \sqrt{\frac{(\xi+c^2)(\eta+c^2)(\zeta+c^2)}{(a^2-c^2)(b^2-c^2)}}
\ee
Let $\Phi_p$ be the potential caused by the ellipsoid,
with the boundary condition $\Phi_p(\xi\rightarrow\infty)=0$.
Requiring continuous boundary condition on the surface of the ellipsoid, we have
\be
\Phi_{\rm in}(0,\eta,\zeta)=\Phi_0(0,\eta,\zeta)+\Phi_p(0,\eta,\zeta).
\label{eq:boundary_E}
\ee
We make the ansatz
\bea
\Phi_p(\xi,\eta,\zeta) & = & F_p(\xi)\sqrt{(\eta+c^2)(\zeta+c^2)},
\eea
which after insertion into the Laplace equation yields
\be
R_\xi\frac{d}{d\xi}\left[R_\xi\frac{dF}{d\xi}\right]-\left(\frac{a^2+b^2}{4}+\frac{\xi}{2}\right)F(\xi)=0.
\ee
Thus, one obtains for the field caused by the ellipsoid
\be
\Phi_p(\xi,\eta,\zeta)=C_pF_p(\xi)\sqrt{(\eta+c^2)(\zeta+c^2)}
\ee
with
\be
F_p(\xi)=F_{\rm in}(\xi)\int\limits_\xi^\infty\frac{d\xi'}{F_{\rm in}^2(\xi')R_{\xi'}},
\ee
where 
\be
F_{\rm in}(\xi)=\sqrt{\xi+c^2},
\ee
the function we used in the case of the charged ellipsoid (see above).
Thus, the field inside the ellipsoid is given by
\be
\Phi_{\rm in}=C_{\rm in}F_{\rm in}(\xi)\sqrt{(\eta+c^2)(\zeta+c^2)}.
\ee

Using the boundary condition shown in Eq.~(\ref{eq:boundary_E}),  one obtains the first equation
\be
C_p\int\limits_0^\infty\frac{d\xi'}{(c^2+\xi')R_{\xi'}}-C_{\rm in}=\frac{E_0}{\sqrt{(a^2-c^2)(b^2-c^2)}},
\ee
The boundary condition of the normal component of $\bD$ at $\xi=0$, equivalent to
\be
\varepsilon_{\rm in}\frac{\p\Phi_{\rm in}}{\p\xi}=\varepsilon_m\frac{\p\Phi_0}{\p\xi}+\varepsilon_m\frac{\p\Phi_p}{\p\xi},
\ee
yields the second equation
\begin{align}
\varepsilon_mC_p\left[\int\limits_0^\infty\frac{d\xi'}{(c^2+\xi')R_{\xi'}}-\frac{2}{abc}\right]-\varepsilon_{\rm in}C_{\rm in} \nn\\
=\frac{\varepsilon_mE_0}{\sqrt{(a^2-c^2)(b^2-c^2)}}.
\end{align}
Consequently, the potentials are
\bea
\Phi_{\rm in} & = & \frac{\Phi_0}{1+\frac{L_3(\varepsilon_{\rm in}-\varepsilon_m)}{\varepsilon_m}}, \\
\Phi_p & = & \Phi_0\frac{\frac{abc}{2}\frac{\varepsilon_m-\varepsilon_{\rm in}}{\varepsilon_m}\int\limits_\xi^\infty\frac{d\xi'}{(c^2+\xi')R_{\xi'}}}{1+\frac{L_3(\varepsilon_{\rm in}-\varepsilon_m)}{\varepsilon_m}},
\eea
where
\be
L_3=\frac{abc}{2}\int\limits_0^\infty\frac{d\xi'}{(c^2+\xi')R_{\xi'}}.
\ee

Far away from the ellipsoid for $\xi\approx r^2\rightarrow\infty$, one can use the approximation
\be
\int\limits_\xi^\infty\frac{d\xi'}{(c^2+\xi')R_{\xi'}}\approx\int\limits_\xi^\infty\frac{d\xi'}{{\xi'}^{5/2}}=\frac{2}{3}\xi^{-3/2},
\ee
yielding the potential caused by the ellipsoid, i.e.
\be
\Phi_p\approx \frac{E_0\cos\theta}{r^2}\frac{\frac{abc}{3}\frac{\varepsilon_{\rm in}-\varepsilon_m}{\varepsilon_m}}{1+\frac{L_3(\varepsilon_{\rm in}-\varepsilon_m)}{\varepsilon_m}},
\ee
from which we identify the dipole moment
\be
\bp=p\hbz=4\pi\varepsilon_m abc\frac{\varepsilon_{\rm in}-\varepsilon_m}{3\varepsilon_m+3L_3(\varepsilon_{\rm in}-\varepsilon_m)}E_0\hbz.
\ee
This result determines the polarizability of the charged ellipsoid, i.e.
\be
\alpha_3=4\pi\varepsilon_m abc\frac{\varepsilon_{\rm in}-\varepsilon_m}{3\varepsilon_m+3L_3(\varepsilon_{\rm in}-\varepsilon_m)}
\ee

If the external electric field is applied along the other major axes of the ellipsoid, $x$ or $y$, the polarizabilities are
\bea
\alpha_1 & = & 4\pi\varepsilon_m abc\frac{\varepsilon_{\rm in}-\varepsilon_m}{3\varepsilon_m+3L_1(\varepsilon_{\rm in}-\varepsilon_m)}, \\
\alpha_2 & = & 4\pi\varepsilon_m abc\frac{\varepsilon_{\rm in}-\varepsilon_m}{3\varepsilon_m+3L_2(\varepsilon_{\rm in}-\varepsilon_m)},
\eea
respectively, where
\bea
L_1 & = & \frac{abc}{2}\int\limits_0^\infty\frac{d\xi'}{(a^2+\xi')R_{\xi'}}, \\
L_2 & = & \frac{abc}{2}\int\limits_0^\infty\frac{d\xi'}{(b^2+\xi')R_{\xi'}}. 
\eea

For oblate spheroids ($a=b$), $L_1=L_2$,
\bea
L_1 & = & \frac{g(e_o)}{2e_o^2}\left[\frac{\pi}{2}-\arctan g(e_o)\right]-\frac{g^2(e_o)}{2}, \nn\\
g(e_o) & = & \sqrt{\frac{1-e_o^2}{e_o^2}},\; e_o^2=1-\frac{c^2}{a^2},
\eea
where $e_o$ is the eccentricity of the oblate spheroid. The limiting cases of an infinitesimally thin disk and a sphere are obtained for $e_o=1$ and $e_o=0$, respectively.

The geometrical factors $L_i$ are related to the depolarization factors $\hat{L}_i$ by
\bea
E_{{\rm in}x} & = & E_{0x}-\hat{L}_1P_{{\rm in}x},\\
E_{{\rm in}y} & = & E_{0y}-\hat{L}_2P_{{\rm in}y},\\
E_{{\rm in}z} & = & E_{0z}-\hat{L}_3P_{{\rm in}z},
\eea
with
\be
\hat{L}_i=\frac{\varepsilon_{\rm in}-\varepsilon_m}{\varepsilon_{\rm in}-\varepsilon_0}\frac{L_i}{\varepsilon_m}.
\ee

\subsection{Dipole Moment of Conducting Ellipsoid induced by an External Electric Field in $x$-direction}
In analogy to Ref.~\onlinecite{Bohren1998}, let us consider the case when the external electric field is parallel to one of the major axes of the ellipsoid,
in this case along the $x$-axis.
For the external potential let us choose
\be
\Phi_0=-E_0 x=-E_0 \sqrt{\frac{(\xi+a^2)(\eta+a^2)(\zeta+a^2)}{(b^2-a^2)(c^2-a^2)}}.
\ee
Let $\Phi_p$ be the potential caused by the ellipsoid,
with the boundary condition $\Phi_p(\xi\rightarrow\infty)=0$.
Requiring continuous boundary condition on the surface of the ellipsoid, we have
\be
\Phi_{\rm in}(0,\eta,\zeta)=\Phi_0(0,\eta,\zeta)+\Phi_p(0,\eta,\zeta).
\label{eq:boundary_Ex}
\ee
Thus, one obtains for the field caused by the ellipsoid
\be
\Phi_p(\xi,\eta,\zeta)=C_pF_p(\xi)\sqrt{(\eta+a^2)(\zeta+a^2)}
\ee
with
\be
F_p(\xi)=F_{\rm in}(\xi)\int\limits_\xi^\infty\frac{d\xi'}{F_{\rm in}^2(\xi')R_{\xi'}},
\ee
where 
\be
F_{\rm in}(\xi)=\sqrt{\xi+a^2},
\ee
the function we used in the case of the charged ellipsoid (see above).
Thus, the field inside the ellipsoid is given by
\be
\Phi_{\rm in}=C_{\rm in}F_{\rm in}(\xi)\sqrt{(\eta+a^2)(\zeta+a^2)}.
\ee

Using the boundary condition shown in Eq.~(\ref{eq:boundary_Ex}),  one obtains the first equation
\be
C_p\int\limits_0^\infty\frac{d\xi'}{(a^2+\xi')R_{\xi'}}-C_{\rm in}=\frac{E_0}{\sqrt{(b^2-a^2)(c^2-a^2)}},
\ee
The boundary condition of the normal component of $\bD$ at $\xi=0$, equivalent to
\be
\varepsilon_{\rm in}\frac{\p\Phi_{\rm in}}{\p\xi}=\varepsilon_m\frac{\p\Phi_0}{\p\xi}+\varepsilon_m\frac{\p\Phi_p}{\p\xi},
\ee
yields the second equation
\begin{align}
\varepsilon_mC_p\left[\int\limits_0^\infty\frac{d\xi'}{(a^2+\xi')R_{\xi'}}-\frac{2}{abc}\right]-\varepsilon_{\rm in}C_{\rm in} \nn\\
=\frac{\varepsilon_mE_0}{\sqrt{(b^2-a^2)(c^2-a^2)}}.
\end{align}
Consequently, the potentials are
\bea
\Phi_{\rm in} & = & \frac{\Phi_0}{1+\frac{L_1(\varepsilon_{\rm in}-\varepsilon_m)}{\varepsilon_m}}, \\
\Phi_p & = & \Phi_0\frac{\frac{abc}{2}\frac{\varepsilon_m-\varepsilon_{\rm in}}{\varepsilon_m}\int\limits_\xi^\infty\frac{d\xi'}{(a^2+\xi')R_{\xi'}}}{1+\frac{L_1(\varepsilon_{\rm in}-\varepsilon_m)}{\varepsilon_m}},
\eea
where
\be
L_1=\frac{abc}{2}\int\limits_0^\infty\frac{d\xi'}{(a^2+\xi')R_{\xi'}}.
\ee

Far away from the ellipsoid for $\xi\approx r^2\rightarrow\infty$, one can use the approximation
\be
\int\limits_\xi^\infty\frac{d\xi'}{(a^2+\xi')R_{\xi'}}\approx\int\limits_\xi^\infty\frac{d\xi'}{{\xi'}^{5/2}}=\frac{2}{3}\xi^{-3/2},
\ee
yielding the potential caused by the ellipsoid, i.e.
\be
\Phi_p\approx \frac{E_0\cos\theta}{r^2}\frac{\frac{abc}{3}\frac{\varepsilon_{\rm in}-\varepsilon_m}{\varepsilon_m}}{1+\frac{L_1(\varepsilon_{\rm in}-\varepsilon_m)}{\varepsilon_m}},
\ee
from which we identify the dipole moment
\be
\bp=p\hbx=4\pi\varepsilon_m abc\frac{\varepsilon_{\rm in}-\varepsilon_m}{3\varepsilon_m+3L_1(\varepsilon_{\rm in}-\varepsilon_m)}E_0\hbx.
\ee
This result determines the polarizability of the charged ellipsoid, i.e.
\be
\alpha_1=4\pi\varepsilon_m abc\frac{\varepsilon_{\rm in}-\varepsilon_m}{3\varepsilon_m+3L_1(\varepsilon_{\rm in}-\varepsilon_m)}
\ee

\subsection{Dipole Moment of Conducting Single-sheet Hyperboloid with a Small Elliptical Wormhole induced by an External Electric Field}
Contrary to the case of an uncharged ellipsoid, where the solutions when applying the external electric field in $x$, $y$, or $z$ direction are similar,
the solutions in the case of an uncharged hyperboloid depend strongly on the axis in which the external field $\bE_0$ points.
While the solutions for $\bE_0=E_0\hbx$ and $\bE_0=E_0\hby$ are similar, the solution for $\bE_0=E_0\hbz$ is completely different.
The reason for this fundamental difference is that the ellipsoid resembles a sphere from far away.
However, a single-sheet hyperboloid has elliptical cylindrical symmetry.

\begin{figure}
\begin{centering}
\includegraphics[width=8.5cm]{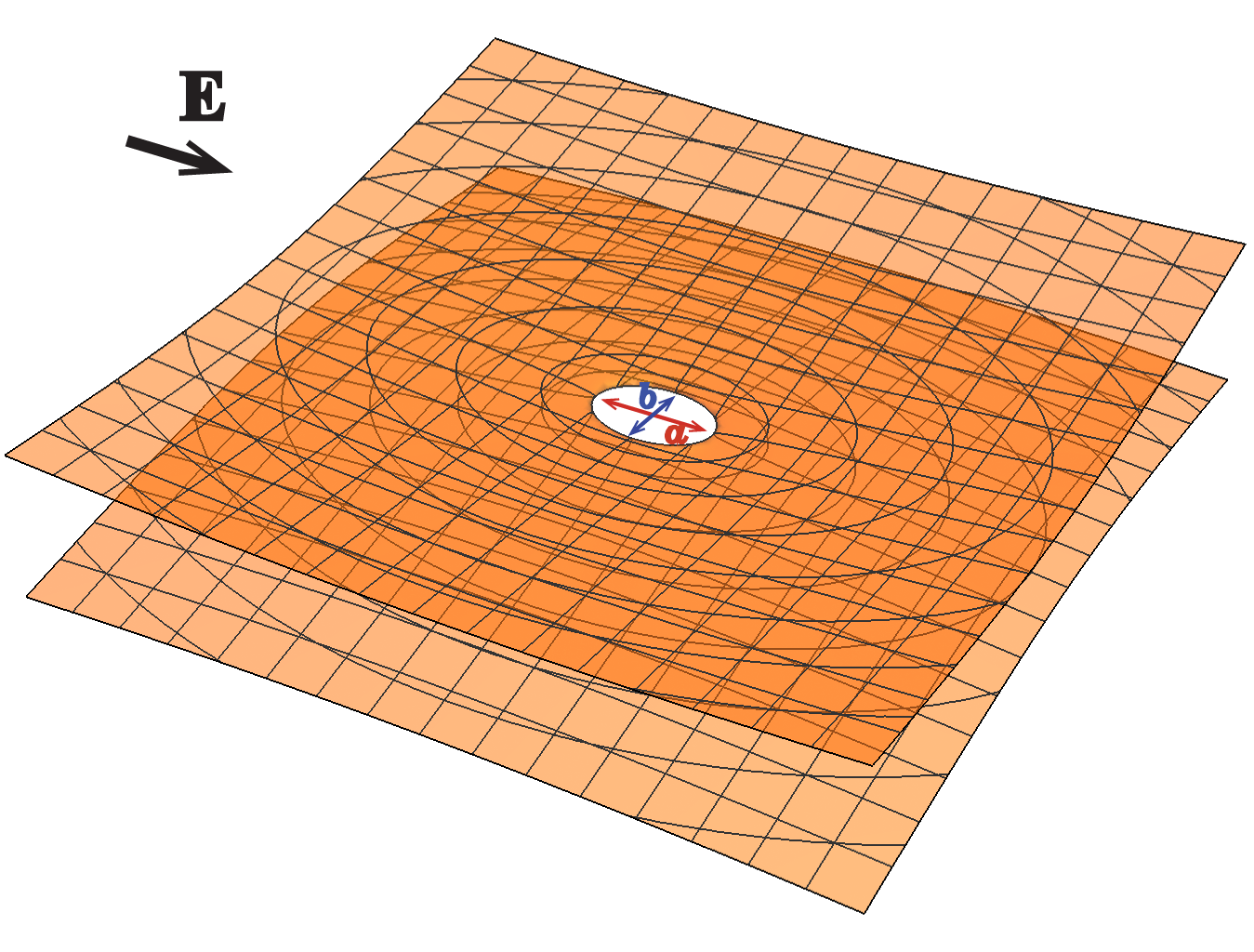}
\end{centering}
\caption{Schematic showing single-sheet hyperboloid with an elliptical wormhole of length $a$, width $b$, and depth $c=0$.
The electric field $\bE_0$ points along the $a$ axis of the ellipse.
\label{fig:wormhole} }
\end{figure}

Here, let us first calculate the electrostatic potential $\Phi(\xi,\eta,\zeta)$ of a conducting single-sheet hyperboloid with an elliptical hole,
which can be represented by a limiting hyperboloid from a family of hyperboloids described by the implicit equation
\be
\frac{x^2}{a^2+u}+\frac{y^2}{b^2+u}+\frac{z^2}{c^2+u}=1
\label{eq:hyperboloid_c0}
\ee
for $a>b>c$. The cubic roots $\xi$, $\eta$, and $\zeta$ are all real in the ranges
\be
-a^2\le\zeta\le -b^2,\; -b^2\le\eta\le -c^2,\; -c^2\le\xi <\infty,
\ee
which are the ellipsoidal coordinates of a point $(x,y,z)$.
The lmiting hyperboloid is a single planar sheet with an elliptical hole, i.e. it belongs to the family of solutions $\eta$ in the limit $\eta\rightarrow -c^2$.
Therefore, let us choose this limiting case as our origin in ellipsoidal coordinates with $c=0$.
Then  Eq.~(\ref{eq:hyperboloid_c0}) becomes
\be
\frac{x^2}{a^2+u}+\frac{y^2}{b^2+u}+\frac{z^2}{u}=1
\ee
for $a>b>c=0$. The cubic roots $\xi$, $\eta$, and $\zeta$ are all real in the ranges
\be
-a^2\le\zeta\le -b^2,\; -b^2\le\eta\le 0,\; 0\le\xi <\infty,
\ee
The surface of the conducting hyperboloid is defined by $-b^2\le\eta=\eta_1<0$.

Let us consider the case $\bE_0=E_0\hbx$, which in the limit when the hyperboloid becomes a flat plane is the most relevant one.
Therefore
\be
\Psi_0=-E_0 x=\mp E_0 \sqrt{\frac{(\xi+a^2)(\eta+a^2)(\zeta+a^2)}{(b^2-a^2)(-a^2)}}
\ee
in the lower-half plane, where the negative sign corresponds to positive $x$ values and the positive sign to negative $x$ values.
Since the equipotential surfaces are determined by $\eta$,
let $\Psi_p$ be the potential caused by the hyperboloid,
with the boundary condition $\Psi_{\rm in}(\eta=0)=0$.
Requiring continuous boundary condition on the surface of the hyperboloid, we have
\bea
\Psi_{\rm in}(\xi,\eta_1,\zeta) & = & \Psi_0(\xi,\eta_1,\zeta)+\Psi_p(\xi,\eta_1,\zeta) , \\
\varepsilon_{\rm in}\left.\frac{\p\Psi_{\rm in}}{\p\eta}\right|_{\eta_1} & = & \varepsilon_m\left.\frac{\p\Psi_0}{\p\eta}\right|_{\eta_1}+\varepsilon_m\left.\frac{\p\Psi_p}{\p\eta}\right|_{\eta_1},
\label{eq:boundary_hyperboloid_Ex}
\eea
where in the second equation the normal component of $\bD$ at $\eta=\eta_1$ must be continuous.
Then we make the ansatz for the electrostatic potential inside the hyperboloid,
\be
\Psi_{\rm in}(\xi,\eta,\zeta)=-C_{\rm in}E_0x,
\ee
where $C_{\rm in}$ is a constant.
This ansatz satisfies the boundary condition
$\Psi_{\rm in}(\xi=0,\eta\rightarrow 0,\zeta)=0$.
For the outside polarization field we choose
\be
\Psi_p(\xi,\eta,\zeta) = -C_pE_0xF_1(\xi)K_1(\eta)
\ee
where $C_p$ is a constant, and we defined
\bea
F_1(\xi) & = & \int\limits_{\xi}^{\infty} \frac{ad\xi'}{2{\xi'}^{1/2}(\xi'+a^2)}- \int\limits_{\xi}^{\infty} \frac{ad\xi'}{2(\xi'+a^2)^{3/2}} \nn\\
& = & \arctan\left(\frac{a}{\sqrt{\xi}}\right)-\frac{a}{\sqrt{\xi+a^2}}.
\eea
Note that $\lim_{\xi\rightarrow 0_+}\arctan\left(\frac{a}{\sqrt{\xi}}\right)=\pi/2$,
whereas $\lim_{\xi\rightarrow 0_-}\arctan\left(\frac{a}{\sqrt{\xi}}\right)=-\pi/2$.
Therefore, in order to avoid discontinuity at $\xi=0$, we must have
$\arctan\left(\frac{a}{-\sqrt{\xi}}\right)=\pi-\arctan\left(\frac{a}{\sqrt{\xi}}\right)$.
\be
K_1(\eta)=\int\limits_{\eta}^\infty \frac{d\eta'}{(\eta'+a^2)R_{\eta'}},
\ee
where $R_\eta = \sqrt{(\eta+a^2)(\eta+b^2)(-\eta)}$.
The boundary conditions at $z\rightarrow\pm\infty$ are satisfied:
\bea
F_1(\xi) =\left\{ \baa
0 & \mbox{ for } z\rightarrow +\infty \\
\pi & \mbox{ for } z\rightarrow -\infty
\ea  \right..
\eea
At large distances $r=\sqrt{x^2+y^2+z^2}$ from the wormhole we have $\xi\approx r^2$.
Then the far-field potential in the upper half-space, which is given by the pure polarization field, is 
\bea
\Psi_p(\xi,\eta,\zeta) & \approx & -C_pE_0x K_1(\eta\approx -b^2)\frac{1}{3}\left(\frac{a}{\sqrt{\xi}}\right)^3 \nn\\
& \approx & -C_pE_0 K_1(\eta\approx -b^2) \frac{a^3}{3}\frac{x}{r^3}.
\eea
The polarization far-field has the form of a dipole field at large distances $r$ from the wormhole.

In order to determine the polarizability of the wormhole, let us find the solution at $\xi=0$, corresponding to the plane that passes through the center of the wormhole.
For $\xi=0$, the unit vectors $\hbx$ and $\hbeta$ are parallel.
In this near-field limit, the polarization potential has the form
\be
\Psi_p(\xi,\eta,\zeta) = -\tC_pE_0x K_1(\eta),
\ee
where $\tC_p=C_p\left(\pi/2-1\right)$.

Using the boundary conditions shown in Eq.~(\ref{eq:boundary_hyperboloid_Ex}),  we obtain the first equation
\be
\tC_pK_1(\eta_1)-C_{\rm in} = 1,
\ee
and the second equation
\begin{align}
& \varepsilon_m\tC_p\left[K_1(\eta_1)\left.\frac{\p x}{\p\eta}\right|_{\eta_1}+K_1'(\eta_1)\left. x\right|_{\eta_1}\right] \nn\\
& -\varepsilon_{\rm in}C_{\rm in}\left.\frac{\p x}{\p\eta}\right|_{\eta_1} =  \varepsilon_m\left.\frac{\p x}{\p\eta}\right|_{\eta_1}.
\end{align}
Using the derivatives
\bea
\left.\frac{\p x}{\p\eta}\right|_{\xi=0,\eta_1} & = &  \frac{a}{2}\sqrt{\frac{(\zeta+a^2)}{(\eta_1+a^2)(a^2-b^2)(a^2-c^2)}}, \\
K_1'(\eta_1) & = & \frac{1}{(\eta_1+a^2)R_{\eta_1}}
\eea
we can rewrite the second equation as
\begin{align}
& \varepsilon_m\tC_p\left[\frac{K_1(\eta_1)}{\eta_1+a^2}+K_1'(\eta_1)\right] \nn\\
& -\varepsilon_{\rm in}C_{\rm in}\frac{1}{\eta_1+a^2} =  \varepsilon_m\frac{1}{\eta_1+a^2},
\end{align}
which is equivalent to
\begin{align}
& \varepsilon_m\tC_p\left[K_1(\eta_1)+\frac{1}{R_{\eta_1}}\right] \nn\\
& -\varepsilon_{\rm in}C_{\rm in} =  \varepsilon_m .
\end{align}
Thus, the potentials are
\bea
\Psi_{\rm in} & = & \frac{\Psi_0}{1+\frac{L_1(\varepsilon_{\rm in}-\varepsilon_m)}{\varepsilon_m}}, \\
\Psi_p & = & \Psi_0\frac{R_{\eta_1}\frac{\varepsilon_m-\varepsilon_{\rm in}}{\varepsilon_m}F_1(\xi)K_1(\eta)(\pi/2-1)}{1+\frac{L_1(\varepsilon_{\rm in}-\varepsilon_m)}{\varepsilon_m}}.
\eea
Then the far-field potential in the upper half-space, which is given by the pure polarization field, is 
\bea
\Psi_p & \approx & -E_0\frac{R_{\eta_1}\frac{\varepsilon_m-\varepsilon_{\rm in}}{\varepsilon_m} 
K_1(\eta\approx -b^2)(\pi/2-1)}{1+\frac{L_1(\varepsilon_{\rm in}-\varepsilon_m)}{\varepsilon_m}}
\frac{a^3}{3}\frac{x}{r^3} \nn\\
& \approx & -E_0\frac{ab\sqrt{-\eta_1}\frac{\varepsilon_m-\varepsilon_{\rm in}}{\varepsilon_m} 
\frac{2\pi}{a^3}(\pi/2-1)}{1+\frac{L_1(\varepsilon_{\rm in}-\varepsilon_m)}{\varepsilon_m}}
\frac{a^3}{3}\frac{x}{r^3} \nn\\
& = & -E_0\frac{ab\sqrt{-\eta_1}\frac{\varepsilon_m-\varepsilon_{\rm in}}{\varepsilon_m} 
\pi(\pi/2-1)}{1+\frac{L_1(\varepsilon_{\rm in}-\varepsilon_m)}{\varepsilon_m}}
\frac{2x}{3r^3},
\eea
where we assumed that $a\approx b$.
The polarization far-field has the form of a dipole field at large distances $r$ from the wormhole.
If the external electric field is applied in $y$-direction, we obtain the potentials
\bea
\Psi_{\rm in} & = & \frac{\Psi_0}{1+\frac{L_2(\varepsilon_{\rm in}-\varepsilon_m)}{\varepsilon_m}}, \\
\Psi_p & = & \Psi_0\frac{R_{\eta_1}\frac{\varepsilon_m-\varepsilon_{\rm in}}{\varepsilon_m}F_2(\xi)K_2(\eta)(\pi-1)}{1+\frac{L_2(\varepsilon_{\rm in}-\varepsilon_m)}{\varepsilon_m}},
\eea
with
\bea
F_2(\xi) & = & \int\limits_{\xi}^{\infty} \frac{bd\xi'}{2{\xi'}^{1/2}(\xi'+b^2)}- \int\limits_{\xi}^{\infty} \frac{bd\xi'}{2(\xi'+b^2)^{3/2}}, \\
K_2(\eta) & = & \int\limits_{\eta}^\infty \frac{d\eta'}{(\eta'+b^2)R_{\eta'}}.
\eea
We defined the geometrical factors
\bea
L_1 & = & R_{\eta_1}K_1(\eta_1)\approx ab\sqrt{-\eta_1}\int\limits_{\eta_1}^\infty \frac{d\eta'}{(\eta'+a^2)R_{\eta'}}, \\
L_2 & = & R_{\eta_1}K_2(\eta_1)\approx ab\sqrt{-\eta_1}\int\limits_{\eta_1}^\infty \frac{d\eta'}{(\eta'+b^2)R_{\eta'}},
\label{eq:L}
\eea
which are related to the depolarization factors by
\be
\tL_i = \frac{\varepsilon_{\rm in}-\varepsilon_m}{\varepsilon_{\rm in}-\varepsilon_0}\frac{L_i}{\varepsilon_m}.
\ee
This result determines the polarizability of the uncharged hyperboloid observable in the far-field, i.e.
\be
\alpha_1= \frac{2ab\sqrt{-\eta_1}\pi(\pi/2-1)}{3}\frac{\varepsilon_{\rm in}-\varepsilon_m}{\varepsilon_m+L_1(\varepsilon_{\rm in}-\varepsilon_m)}.
\label{eq:polarizability_hyperboloidx}
\ee
Similarly, we obtain the polarizability in $y$-direction, i.e.
\be
\alpha_2 = \frac{2ab\sqrt{-\eta_1}\pi(\pi/2-1)}{3}\frac{\varepsilon_{\rm in}-\varepsilon_m}{\varepsilon_m+L_2(\varepsilon_{\rm in}-\varepsilon_m)}.
\ee
Comparing to the polarizabilities of ellipsoids,\cite{Bohren1998} the polarizabilities of hyperboloids are proportional to $ab\sqrt{-\eta_1}$, which corresponds to the volume of the ellipsoid $abc$.

In the case of circular wormholes, we have $a=b$, and therefore $\alpha_1=\alpha_2=\alpha_{||}$,
with $L_1=L_2=L_{||}$.

\subsection{Dispersion relations}
In our proposed mid-IR light source the effective combination of Si$_3$N$_4$ and h-BN behaves as an environment with polar phonons. 
Polar materials have ions of different valence, whose oscillating dipole moment gives rise to the interaction between electrons and optical phonons, the Frohlich interaction.
By placing graphene on a polar substrate the long range Frohlich interaction mediates the interaction between optical phonons and surface plasmons in graphene.\cite{Yan2013}
The interaction between polar substrate/graphene phonons and electrons in graphene modifies substantially the graphene plasmon dispersion relation, which is shown in Fig.~\ref{fig:dispersion}. 
The dielectric function of graphene in the random phase approximation (RPA) takes the form\cite{Safaei2017,Paudel2017}
\bea
{{\bf{\varepsilon }}^{RPA}}({\bf{q}},\omega ) & = & {{\bf{\varepsilon }}_m} - {v_c}({\bf{q}}){\chi ^0}({\bf{q}},\omega )  \nn\\
& & - {{\bf{\varepsilon }}_m}\sum\limits_l {{v_{sph,l}}({\bf{q}},\omega )} {\chi ^0}({\bf{q}},\omega ) \nn\\
& &   - {{\bf{\varepsilon }}_m}{v_{oph}}({\bf{q}},\omega )\chi _{j,j}^0({\bf{q}},\omega ).
\label{eq:RPA}
\eea
The second term represents the effective Coulomb interaction of electrons in graphene, and 
${v_c}({\bf{q}}) = {{{e^2}} \mathord{\left/
 {\vphantom {{{e^2}} {2q}}} \right.
 \kern-\nulldelimiterspace} {2q}}{\varepsilon _0}$  is the direct Coulomb interaction. The third term is the effective dielectric function for different phonon modes ($l$) coming from electron-electron interaction mediated by substrate optical phonons, which couple to the electrons by means of the Frohlich interaction, i.e.
 \be
{v_{sph,l}}({\bf{q}},\omega ) = {\left| {{M_{sph}}} \right|^2}G_l^0(\omega ),
 \ee
 where $|M_{sph} |^2$ is the scattering and
$G_l^0$ is the free phonon Green function. The last term of Eq.~(\ref{eq:RPA}) corresponds to the optical phonon mediated electron-electron interaction
\be
{v_{oph}}({\bf{q}},\omega ) = {\left| {{M_{oph}}} \right|^2}{G^0}(\omega ).
\ee
Here $|M_oph |^2$ defines the scattering matrix element and ${G^o}(\omega )$ is the free phonon Green function. In Eq.~(\ref{eq:RPA}), 
$\chi _{j,j}^0({\bf{q}},\omega )$ is the current-current correlation function. This description is very general and can be applied to any metallic system.

\begin{figure}
\begin{centering}
\includegraphics[width=8.5cm]{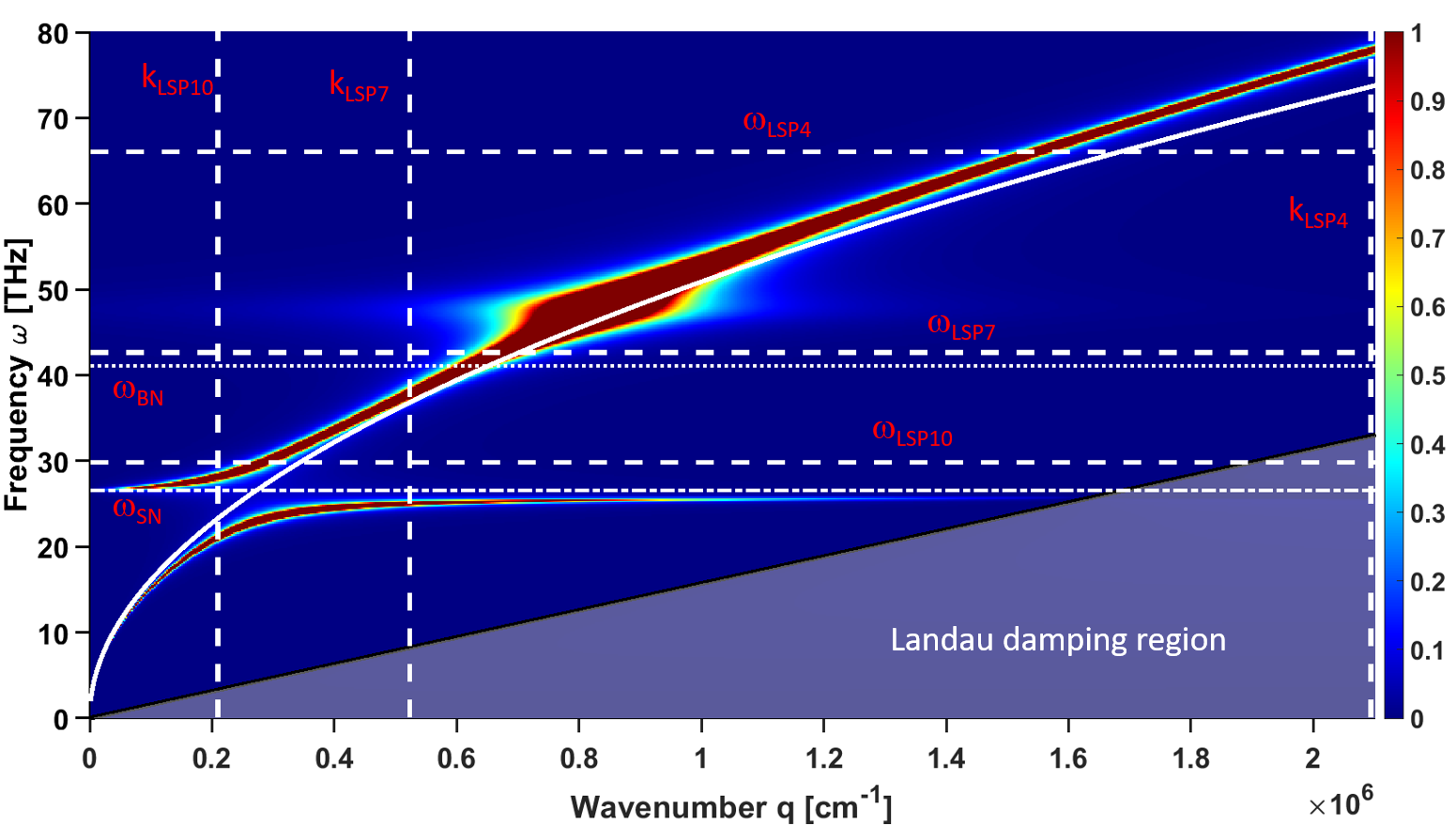}
\end{centering}
\caption{The energy loss function for graphene with  $\epsilon_F=-1.0$ eV. $k_{LSP4}$, $k_{LSP7}$, and $k_{LSP10}$  are the plasmon wavenumber associated with the nanopatterning of the graphene sheet shown in Figs.~\ref{fig:spectral_radiance_4m}, \ref{fig:spectral_radiance_7m}, and \ref{fig:spectral_radiance_10m}, respectively.  $\omega_{LSP4}$, $\omega_{LSP7}$, and $\omega_{LSP10}$  represent the LSP resonances shown in Figs.~\ref{fig:spectral_radiance_4m}, \ref{fig:spectral_radiance_7m}, and \ref{fig:spectral_radiance_10m}, respectively. The polar phonon resonance of h-BN and the surface polar phonon resonance of Si$_3$N$_4$ are denoted by $\omega_{BN}$, and $\omega_{SN}$, respectively. The Landau damping region is marked by the shaded area.
\label{fig:dispersion} }
\end{figure}

The momentum relaxation time $\tau$ can be derived by considering the impurity, electron-phonon interaction, and the scattering related to nanostructure edges 
$\tau^{-1}=\tau_{DC}^{-1}+\tau_{edge}^{-1}+\tau_{e-p}^{-1}$,
which determines the plasmon lifetime and the absorption spectrum bandwidth. It can be evaluated via the measured DC mobility $\mu$  of the graphene sample through 
${\tau _{DC}} = {{\mu \hbar \sqrt {\pi \rho } } \mathord{\left/
 {\vphantom {{\mu \hbar \sqrt {\pi \rho } } {e{v_F}}}} \right.
 \kern-\nulldelimiterspace} {e{v_F}}}$,
 where $v_F=10^6$ m/s is the Fermi velocity and
 ${\tau _{DC}} = {{\mu \hbar \sqrt {\pi \rho } } \mathord{\left/
 {\vphantom {{\mu \hbar \sqrt {\pi \rho } } {e{v_F}}}} \right.
 \kern-\nulldelimiterspace} {e{v_F}}}$
 is the charge carrier density. 
 ${\tau _{edge}} \approx {\left( {{{1 \times {{10}^6}({m \mathord{\left/
 {\vphantom {m s}} \right.
 \kern-\nulldelimiterspace} s})} \mathord{\left/
 {\vphantom {{1 \times {{10}^6}({m \mathord{\left/
 {\vphantom {m s}} \right.
 \kern-\nulldelimiterspace} s})} {w - {w_0}}}} \right.
 \kern-\nulldelimiterspace} {w - {w_0}}}} \right)^{ - 1}}$
is due to the scattering from the nanostructure edges, where $w$ is the edge-to-edge distance of the holes and 
$w_0=7$ nm is the parameter that includes edge effects, and
${\tau _{e - ph}} = {\hbar  \mathord{\left/
 {\vphantom {\hbar  {2{\mathop{\rm Im}\nolimits} ({\sum _{e - ph}})}}} \right.
 \kern-\nulldelimiterspace} {2{\mathop{\rm Im}\nolimits} ({\sum _{e - ph}})}}$
 is related to the scattering because of coupling of electrons and phonons.
 The imaginary part of the electron-phonon self-energy is given by
 \be
{\mathop{\rm Im}\nolimits} ({\sum _{e - ph}}) = \gamma \left| {\hbar \omega  - {\mathop{\rm sgn}} \left( {\hbar \omega  - {E_F}} \right)\hbar {\omega _{oph}}} \right|.
 \ee
 $\gamma  = 18.3 \times {10^{ - 3}}$ is a dimensionless constant describing the electron-phonon coupling coefficient, and 
$\hbar {\omega _{oph}} \approx 0.2$ eV is the graphene optical phonon energy. 
It is evident that the plasmon lifetime is reduced due to electron-phonon interaction and edge scattering, but the DC conductivity, which is used to calculate the dielectric function of graphene, is invariant if the edge-to-edge distance of the pattern is larger than the carrier mean free path ${L_{MFP}} = {v_F}{\tau _{DC}}$.

The Drude model and the optical phonon frequency isare not valid for a patterned graphene sheet only if the edge-to-edge distance is much lowersmaller than the carrier mean free paths of electrons and phonons.  For the current pattern and carrier mobilities, both the mean free paths $L_e$ and $L_{ph}$ are smaller than or of the same order as the edge-to-edge distance, which means that we can safely assume that the Drude model and optical phonon frequencies of our patterned graphene sheet are the same as for pristine graphene.
The coupling of plasmon and substrate/graphene phonon can be characterized through the loss function $Z$, which is the imaginary part of inverse effective dielectric function calculated via the generalized RPA theory 
\be
Z \propto  - {\mathop{\rm Im}\nolimits} \left( {\frac{1}{{{\varepsilon ^{RPA}}}}} \right).
\ee
The loss function represents the amount of energy dissipated by exciting the plasmon coupled to the substrate and optical phonons in graphene. The surface plasmons in graphene are damped through radiative and nonradiative processes.
Nonradiative damping transfers the plasmon energy to hot electron-hole excitation by means of intraband transition. 
Fig.~\ref{fig:dispersion} (a) shows the loss function for graphene with carrier mobility $\mu=3000$ cm$^2$/V$\cdot$s and a Fermi energy of $\epsilon_F=1.0$ eV.
The thickness of the optical cavity is chosen to be $\lambda/4n$, where $n$ is the refractive index of the cavity material.\cite{Safaei2017}
The plasmon assisted electron-hole pair generation in this structure lies outside the Landau intraband damping region, indicated by the shaded area in Fig.~\ref{fig:dispersion} (a).
A band gap in the plasmon-phonon dispersion relation is formed via Frohlich interaction between graphene plasmons and optical phonons.
This coupling leads to the splitting of the energy into two distinct branches: surface plasmon phonon polaritons (SPPPs) and graphene plasmons (GPs).
The horizontal branch line marked as $\omega_{LSP}$ is the LSP mode and is independent of the plasmon wavevector due to the localization of the LSP around a hole.
The resonance frequencies of the polar phonons are denoted by $\omega_{BN}$ for h-BN and by $\omega_{SN}$ for Si$_3$N$_4$.

\subsection{Integral of dyadic Green function elements over spherical angle}
For the calculation of the spectral radiance we need to integrate the elements of the dyadic Green function over the spherical angle.
We can split the total dyadic Green function into a free space term $\obG_0(\br,\br';\omega)$ and a term $\obG_{SPP}(\br,\br';\omega)$ that creates surface plasmon polaritons inside graphene.
Since the absorbance of the pristine graphene sheet is only 2.3\%, we can safely neglect $\obG_{SPP}(\br,\br';\omega)$. Our goal is to calculate the gray-body emission of the EM radiation from the LSP around the holes in graphene into free space. Therefore, we need to evaluate
\be
I_{GB}^\infty(\omega)=\lim_{r\rightarrow\infty}\int r^2\sin\theta d\theta d\varphi I_{GB}(r,\omega),
\ee
where can use the approximation
\be
I_{GB}(r,\omega)=I_0(r,\omega)-I_{SPP}(r,\omega)\approx I_0(r,\omega).
\ee
In Cartesian coordinates, we can write down the dyadic Green function as\cite{Novotny2012}
\bea
\obG_0(\br;\omega) & = & \frac{e^{ikr}}{4\pi r}\left[\left(1+\frac{i}{kr}-\frac{1}{k^2r^2}\right)\unit \right. \nn\\
& & +\left.\left(\frac{3}{k^2r^2}-\frac{3i}{kr}-1\right)\hbr\hbr\right].
\eea
Since we are interested only in the far field, we consider only the far-field component of the dyadic Green function, which is
\bea
\obG_{FF}(\br;\omega) = \frac{e^{ikr}}{4\pi r}\left[\unit- \hbr\hbr \right] ,
\eea
which possesses only angular (transverse) components but no radial (longitudinal) components.
Then the necessary components are
\bea
G_{xx}(\br;\omega) & = & \frac{e^{ikr}}{4\pi r}\left[1-\sin^2\theta\cos^2\varphi\right], \nn\\
G_{yx}(\br;\omega) & = & \frac{e^{ikr}}{4\pi r}\left[1-\sin^2\theta\cos\varphi\sin\varphi\right] \nn\\
G_{zx}(\br;\omega) & = & \frac{e^{ikr}}{4\pi r}\left[1-\sin\theta\cos\theta\cos\varphi\right], \nn\\
G_{xy}(\br;\omega) & = & \frac{e^{ikr}}{4\pi r}\left[1-\sin^2\theta\cos\varphi\sin\varphi\right], \nn\\
G_{yy}(\br;\omega) & = & \frac{e^{ikr}}{4\pi r}\left[1-\sin^2\theta\sin^2\varphi\right], \nn\\
G_{zy}(\br;\omega) & = & \frac{e^{ikr}}{4\pi r}\left[1-\sin\theta\cos\theta\sin\varphi\right], 
\eea
The corresponding integrals are
\bea
\int r^2\sin\theta d\theta d\varphi \left|G_{xx}(\br;\omega)\right|^2 & = & \frac{2}{15\pi}, \nn\\
\int r^2\sin\theta d\theta d\varphi  \left|G_{yx}(\br;\omega)\right|^2 & = & \frac{4}{15\pi}, \nn\\
\int r^2\sin\theta d\theta d\varphi  \left|G_{zx}(\br;\omega)\right|^2 & = & \frac{4}{15\pi}, \nn\\
\int r^2\sin\theta d\theta d\varphi \left|G_{xy}(\br;\omega)\right|^2 & = & \frac{4}{15\pi}, \nn\\
\int r^2\sin\theta d\theta d\varphi  \left|G_{yy}(\br;\omega)\right|^2 & = & \frac{2}{15\pi}, \nn\\
\int r^2\sin\theta d\theta d\varphi  \left|G_{zy}(\br;\omega)\right|^2 & = & \frac{4}{15\pi}.
\eea

\bibliographystyle{apsrev4-1}

%

\end{document}